# On the parametrization of epidemiologic models – lessons from modelling COVID-19 epidemic

Yuri Kheifetz[1], Holger Kirsten[1], Markus Scholz[1]

[1]Institute for Medical Informatics, Statistics and Epidemiology, University of Leipzig, Leipzig, Germany

Corresponding author:

Prof. Dr. Markus Scholz

Institute for Medical Informatics, Statistics and Epidemiology

University of Leipzig

Härtelstraße 16-18

04107 Leipzig

Germany

Email: markus.scholz@imise.uni-leipzig.de

## Abstract

A plethora of prediction models of SARS-CoV-2 pandemic were proposed in the past. Prediction performances not only depend on the structure and features of the model, but also on its parametrization. Official databases are often biased due to lag in reporting of cases, changing testing policy or incompleteness of data. Moreover, model parametrization is time-dependent e.g. due to changing age-structures, new emerging virus variants, non-pharmaceutical interventions and ongoing vaccination programs.

To cover these aspects, we develop a principled approach to parametrize SIR-type epidemiologic models of different complexities by embedding the model structure as a hidden layer into a general Input-Output Non-Linear Dynamical System (IO-NLDS). Non-explicitly modelled impacts on the system are imposed as inputs of the system. Observable data are coupled to hidden states of the model by appropriate data models considering possible biases of the data.

We estimate model parameters including their time-dependence by a Bayesian knowledge synthesis process considering parameter ranges derived from external studies as prior information. We applied this approach on a SIR-type model and data of Germany and Saxony demonstrating good prediction performances.

By our approach, we can estimate and compare for example the relative effectiveness of non-pharmaceutical interventions and can provide scenarios of the future course of the epidemic under specified conditions. Our method of parameter estimation can be translated to other data sets, i.e. other countries and other SIR-type models even for other disease contexts.

# Introduction

Predicting the spread of an infectious disease is a pressing need as demonstrated for the present SARS-CoV-2 pandemic. Due to the worldwide high disease burden, a plethora of mathematical epidemiologic models, often of the SIR type, were proposed and published [1]. Major aims are to predict **(1)** the dynamics of infected subjects, **(2)** requirements of medical resources during the course of the epidemic or **(3)** the effectiveness of non-pharmaceutical intervention programs (NPI) [2] [3] .

A good prediction performance does not only depend on the precise structure of the model but on its parametrization. This, however, is a non-trivial and often underestimated task due to the following issues applicable to other infectious diseases as well: Key epidemiologic parameters are often unknown or only known within ranges. Therefore, parametrization on the basis of observational data is a common approach. However, reported official data bases are heterogeneous and often biased due to **(1)** Lag in reporting of cases / events [4], **(2)** Changing testing policy either due to limited testing capacities, which might depend on the pandemic situation itself or by changing risk profiles of people to be tested (e.g. defined risk groups, dependence on symptoms, degree of prophylactic testing) [5], and, **(3)** incompleteness of data [6]. Moreover, parametrization depends on further epidemiologic issues to be considered, comprising **(1)** changing age-structure of the infected population with impact on symptomatology, hospital or intensive care requirements and mortality, **(2)** spatial heterogeneity of the spread of the disease driven by local conditions and outbreaks, **(3)** new pathogen variants becoming prevalent, **(4)** non-pharmaceutical interventions continuously updated in response to the pandemic situation, and finally, **(5)** the progress of vaccination programs and its effectiveness.

Due to these complexities, which are also interacting, it is close to impossible to construct a fully mechanistic model covering all these aspects in parallel. Therefore, we here propose a framework of epidemiologic model parametrization, which accounts for these issues in a more phenomenological, data-driven way applicable even for limited or biased data resources.

In detail, we here propose to integrate mechanistic epidemiologic models as hidden layers into Input-Output Non-Linear Dynamical Systems (IO-NLDS), i.e. the true epidemiologic dynamics cannot directly be observed. This allows distinguishing between features explicitly modelled (in our case different virus variants, vaccination) and changing factors of the system which are difficult to model mechanistically (in our case changes of contacts, e.g. due to non-pharmaceutical interventions or changing contact behaviour, changing age-structure of the infected population and changes in testing policy, in the following abbreviated as NPI/behavior). These factors are imposed as external inputs of the system.

We then estimate parameters by a knowledge synthesis process considering prior information of parameter ranges derived from different external studies and other available data resources such as public data. Specifically, we use Bayesian inference for the parameter estimation, which could also be time-dependent. We analyze the structure of available public data in detail and translate it to model outputs linked by an appropriate data model to the hidden states of the IO-NLDS, i.e. the epidemiologic model. We demonstrate this approach on an example of our epidemiologic model of the SECIR type for SARS-CoV-2 and data of Germany and Saxony, but our method can be translated to other countries, other models or even other infectious disease contexts.

# Materials and methods

## General approach

We consider Input-Output Non-Linear Dynamical Systems (IO-NLDS) originally proposed as time-discrete alternatives to physiological pharmacokinetic and –dynamics differential equations models [7]. This class of models couples a set of input parameters such as external influences and factors with a set of output parameters, i.e. observations by a hidden model structure to be learned (named *core*

*model* in the following). This coupling is not necessarily fully deterministic, i.e. data are not required do directly represent state parameters of the model. This represents a major feature of our approach in order to account for different types of biases in available observational time series data.

We here demonstrate our approach by using an epidemiological model as core of the IO-NLDS. Non-pharmaceutical interventions, changes of testing policy, age distributions and severity of diseae courses were phenomenologically modelled by external control parameters imposed on the epidemiologic model via the input layer of the IO-NLDS. Random influxes of infected subjects e.g. by travelling activities or outbreaks are also considered by this approach. Number of reported infections, intensive care (IC) capacity and deaths are considered as output parameters not directly representing the hidden states of the model due to several data issues including reporting delays. The model is then fitted to data by a full information approach, i.e. all data points were evaluated by a suitable likelihood function.

The single steps of this process are explained in detail below.

### Assumptions of the core model

We adapted a standard SECIR model (SECIR = Suszeptible, Exposed, Cases, Infectious, Recovered) for pandemic spread. We introduced an asymptomatic compartment in order to account for infected patients, which do not have symptoms, a common condition of SARS-CoV-2 infection. A compartment of patients requiring intensive care (IC) was added to model respective requirements and we distinguished between deceased and recovered patients.

We subdivided most of the compartments into 3 sub-compartments with first order transitions to model time delays. To allow for two concurrent virus variants with differing properties, compartments of asymptomatic and symptomatic infected subjects are duplicated. This allows us, for example, to simulate the take-over of the more infectious B.1.1.7 (Alpha), and later, B1.617.2 (Delta) variant observed e.g. in all European countries [8].

The general scheme of the IO-NLDS system is shown in **Figure 1**. We make the following assumptions:

1. The input layer consists of external modifiers influencing (**1**) reporting policy (e.g. changing testing policy), (**2**) rates of infections (affected by non-pharmaceutical interventions, age structure, influx of cases) and (**3**) risks of severe disease conditions such as IC requirements and deaths, again affected by changing age structure of infected subjects.
2. The output layer of observable data is linked to the hidden layers of the core model by specific data models (see later).
3. Susceptible, non-infected people (*Sc*): We assume that 100% of the population is susceptible to infection in the beginning of the epidemic.
4. The latent state *E* comprises infected but non-infectious people.
5. The asymptomatic infected state $I_A$ has three sub-compartments ($I_{A,1}$, $I_{A,2}$ and $I_{A,3}$). From $I_{A,1}$, transitions to the symptomatic state or the second asymptomatic state are possible. From $I_{A,2}$, only transitions to $I_{A,3}$ and then to the recovered state *R* are assumed.
6. The symptomatic infected state $I_S$ is also divided into three compartments ($I_{S,1}$, $I_{S,2}$ and $I_{S,3}$). The sub-compartment $I_{S,1}$ comprises an efflux towards the sub-compartment $C_1$ representing deteriorations towards critical disease states. Otherwise, the patient transits to $I_{S,2}$. From $I_{S,2}$, a patient can either die representing deaths without prior intensive care or transit to $I_{S,3}$. Finally, the efflux of $I_{S,3}$ flows into *R* representing resolved disease courses.
7. Both, cases in $I_A$ and $I_S$ contribute to new infections but with different rates to account for differences in infectivity and quarantine probabilities.

8. The compartment *C* represents critical disease states requiring intensive care. We assume that these patients are not infectious due to isolation. Again, the compartment is divided into three sub-compartments, $C_1$, $C_2$ and $C_3$. In $C_1$, a patient can either die or transit to $C_2$, $C_3$, and finally, *R*.
9. Patients on the recovered stage *R* are assumed to be immune against re-infections.
10. We duplicate the compartments $E$, $I_{A,1}$,…, $I_{A,3}$, $I_{S,1}$, …, $I_{S,3}$ to account for two concurrent virus variants. We assume different infectivities for the two variants. All other parameters are assumed equal. No co-infections are assumed.

These assumptions are translated into a difference equation system (see Supplemental information S1). Model compartments and their properties are explained in **Table 1**.

*Table 1: Description of model compartments. We describe the compartments of the model and their biological meaning. Compartments E, $I_A$ and $I_S$ are duplicated to account for two concurrent virus variants.*

| Compartment name | Sub-compartments | Description |
|---|---|---|
| $Sc$ | | Susceptible |
| $E$ | | Latent stage (not infectious) |
| $I_A$ | $I_{A,1}$ | Asymptomatic infected state 1. Can either develop symptoms, i.e. transit to $I_{S,1}$ with probability $p_{symp}$ and rate $r_{4b}$ or stays asymptomatic with probability $1 - p_{symp}$ and transits to $I_{A,2}$ with rate $r_4$ |
| | $I_{A,2}$ | Asymptomatic infected state 2, transits to $I_{A,3}$ with rate $r_4$ |
| | $I_{A,3}$ | Asymptomatic infected stage 3 transits to *R* with rate $r_4$ |
| $I_S$ | $I_{S,1}$ | Symptomatic infected state 1. Can either become critical, i.e. transits to $C_1$ with probability $p_{crit}$ and rate $r_6$ or stays sub-critical with probability $1 - p_{crit}$ and transits to $I_{S,2}$ with rate $r_5$ |
| | $I_{S,2}$ | Symptomatic infected state 2, transits to $I_{S,3}$ with rate $r_5$ |
| | $I_{S,3}$ | Symptomatic infected state 3, transits to *R* with rate $r_5$ |
| $C$ | $C_1$ | Critical state 1, not infectious. Can either die, i.e. transits to *D* with probability $p_{death}$ and transit rate $r_8$ or stays critical with probability $1 - p_{death}$ and transits to $I_{S,2}$ with rate $r_7$ |
| | $C_2$ | Critical state 2, transits to $C_3$ with rate $r_7$ |
| | $C_3$ | Critical state 3, transits to *R* with rate $r_7$ |
| $R$ | | Recovered (absorbing state) |
| $D$ | | Dead (absorbing state) |

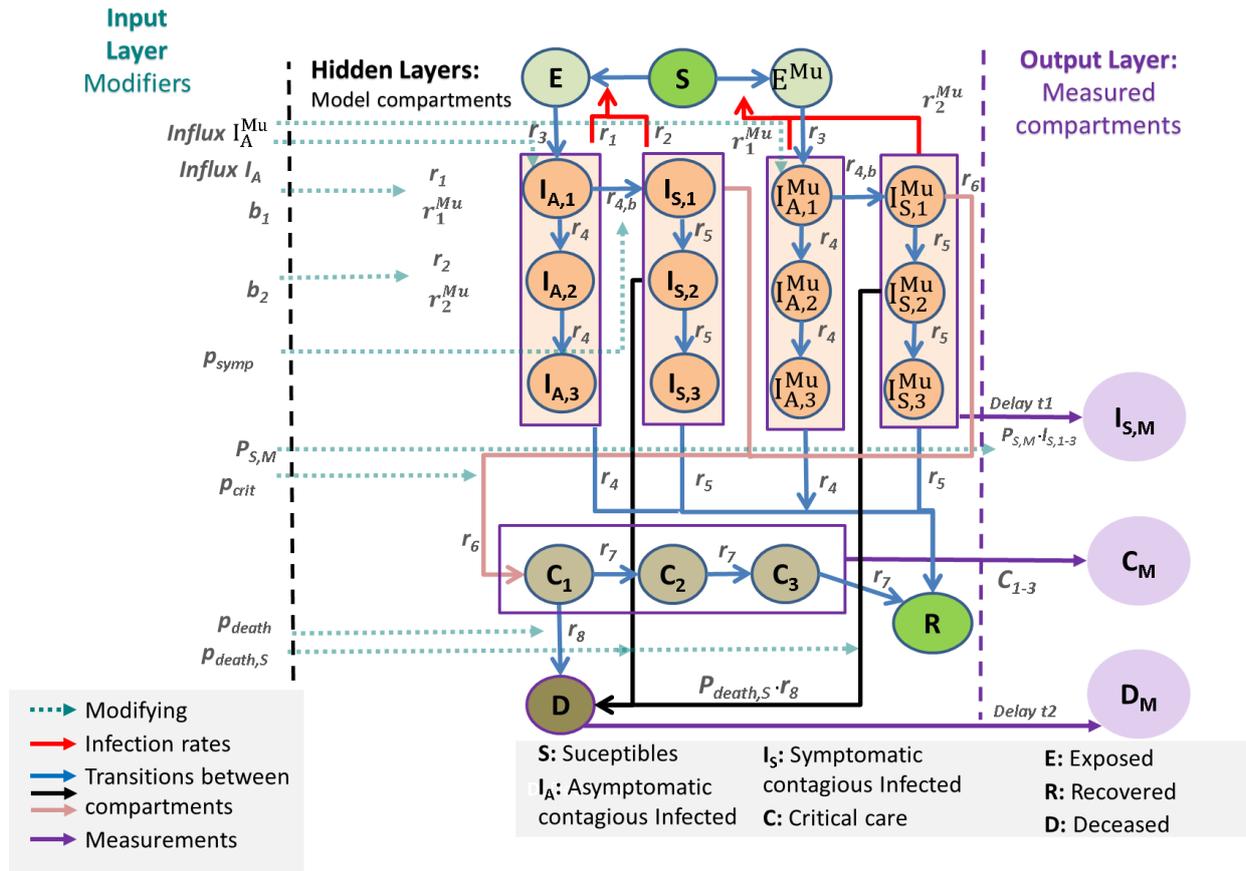

*Figure 1: General scheme of our IO-NLDS model. The epidemiologic SECIR model is integrated as a hidden layer. Respective equations are provided in supplemental information S1. The input layer consists of external modifiers including parameter changes due to changes in testing policy, non-pharamceutical interventions and age-structures. The output layer is derived from respective hidden layers via a stochastic relationships (see later). The output layer is compared with real-world data.*

All model parameters of the model are described in the **Table 2**.

*Table 2: Basic model parameters. We present prior values and ranges derived from the literature as well as estimated values derived from parameter fitting. Transit rate means reverse of transit time of the respective compartment. Further details and definitions on parameters are given in the Supplemental information, e.g. derivation of priors. Posteriors can be found in **Figure 4** and **Supplemental Table S8**.*

| Parameter | Unit | Description | Source | Value | Prior mean | Min | Max |
|---|---|---|---|---|---|---|---|
| influx | Subjects per day | Initial influx of infections into compartment $E$ until first interventions | Estimated | 3171 | - | - | - |
| $r_1$ | Day$^{-1}$ | Infection rate through asymptomatic subjects | Estimated | 1.19 | - | - | - |
| $r_2$ | Day$^{-1}$ | Infection rate through symptomatic subjects | Set equal to $rb_{1,2} \cdot r_1$ (parsimony) | 0.451 | - | - | - |
| $rb_{1,2}$ | - | Proportion of infection rate symptomatics/asymptomatics $r_1/r_2$ | Estimated | 0.379 | - | 0 | - |
| $r_3$ | Day$^{-1}$ | Transit rate for compartment E (latent time) | prior constrain | 0.272 | 1/3 | 1/4 | 1/2 |
| $r_4$ | Day$^{-1}$ | Transit rate for asymptomatic sub-compartments | prior constrain | 0.636 | 3/5 | 3/10 | 3/4 |
| $r_{4,b}$ | Day$^{-1}$ | Rate of development of symptoms after infection | prior constrain | 0.456 | 1/2.5 | 1/5 | 1 |
| $r_5$ | Day$^{-1}$ | Transit rate for symptomatic sub-compartments | prior constrain | 0.946 | 3/2.5 | 3/7.5 | 3/1.5 |
| $r_6$ | Day$^{-1}$ | Rate of development of critical state after being symptomatic | prior constrain | 0.186 | 1/5 | 1/7 | 1/4 |
| $r_7$ | Day$^{-1}$ | Transit rate for critical state sub-compartment | prior constrain | 0.159 | 3/17 | 3/35 | 3/8 |
| $r_8$ | Day$^{-1}$ | Death rate of patients in critical sub-compartment 1 | prior constrain | 0.104 | 1/8 | 1/14 | 1/6.5 |
| $p_{symp}$ | - | Probability of symptoms development after being infected | Set or prior constrained (overfitted if estimated unconstrained) | 0.5 | - | 0.3 | 0.8 |
| $p_{crit}$ ($p_{crit,0}$) | - | Initial value $p_{crit,0}$ of step function $p_{crit}$, the probability of becoming critical after developing symptoms | Estimated | 0.0765 | - | 0 | 1 |
| $p_{death}$ ($p_{death,0}$) | - | Initial value $p_{death,0}$ of step function $p_{crit}$, the probability of becoming critical after developing symptoms | Estimated | 0.119 | - | 0 | 1 |
| $p_{death,S}$ | - | Probability of death after developing symptoms without becoming critical | Set equal to $p_{death,S,0} \cdot p_{death}$ (parsimony) | - | - | 0 | 1 |
| $p_{death,S,0}$ | | Proportionality factor for probability of death after developing symptoms without becoming critical | Estimated | 0.587 | | | |
| $P_{S,M}$ | - | Fraction of unreported cases | prior constrain | 0.499 | 0.5 | 0.1 | 0.90 |
| mur | | Ratio of r1$^{Mu}$/r1 = r2$^{Mu}$/r2 reflecting higher infectivity of B.1.1.7 variant | Set | 1.65 | - | - | - |

Input layer

The input layer represents external factors acting at the SECIR model, effectively changing its parameters. Therefore, we define step functions $b_1$ and $b_2$ as time-dependent input parameters modifying the rate of infections caused by asymptomatic, respectively symptomatic subjects. To identify dates of change, we used a data-driven approach on the basis of a Bayesian Information Criterion informed by changes in non-pharmaceutical interventions for Germany based on Government decisions, changing testing policies as well as events with impact on epidemiological dynamics such as holidays or sudden outbreaks. Details can be found in **Supplemental information S2**.

We also accounted for changes in the probabilities of critical courses and mortality, which can be explained by changes in testing policies covering asymptomatic cases to a different extent (for example symptomatic testing only vs. introduction of screening tests, e.g. rapid antigen tests), respectively shifts in the age-distribution of patients or changes in patient care efficacy (new pharmaceutical treatment, overstretched medical resources). Again, this is implemented by step functions $p_{crit}$, respectively $p_{death}$. Number of steps are determined on the basis of a Bayesian Information Criterion. Details can be found in the **Supplemental information S2** as well as in **Supplemental Table S2**. The parameter $P_{S,M}$ represents the percentage of reported infected symptomatic subjects in relation to all symptomatic subjects. This value is assumed to be constant (50%) in the present version of the model.

*Table 3: Parameters used to define the input layer. These parameters were used to empirically model changing NPIs or changing contact behaviour, changes in testing policies and changing age-structures during the course of the epidemic. Respective input functions constitute the input layer of our IO-NLDS model.*

| wParameter | Unit | Description | Source | Remarks |
|---|---|---|---|---|
| $N_{tr}$ | - | Number of time points of changes of NPI / contact behaviour | Empirically defined | 13 intensifications, 15 relaxations identified (determined by information criterion) |
| $b_{tr,j}$, j=1,…, $N_{tr}$ | - | Relative infectivity of subjects in the time interval [$tr$, $tr+1$] | Estimated | assumed to be the same for symptomatic and asymptomatic patients |
| $Tr_j$, j=1,…, $N_{tr}$ | Days | Time points of NPI / contact behaviour changes | Estimated or fixed | Strictly monotone sequence |
| $N_{crit}$ | - | Number of time steps of $p_{crit}(t)$ | Empirically defined | 18 (determined by information criterion) |
| $\alpha_{crit,j}$, j=1,…, $N_{crit}$ | - | Value of $p_{crit}$ between two time steps | Estimated | [0,1] |
| $T_{pcrit,j}$, j=1,…, $N_{crit}$ | Days | Time points of changes of $p_{crit}$ | Estimated | Strictly monotone sequence |
| $N_{death}$ | - | Number of time steps of $p_{death}(t)$ | Empirically defined | 19 (determined by information criterion) |
| $\alpha_{death,j}$, j=1,…, $N_{death}$ | - | Value of $p_{death}$ between two time steps | Estimated | [0,1] |
| $T_{pdeath,j}$, j=1,…, $N_{death}$ | Days | Time points of changes of $p_{death}$ | Estimated | Strictly monotone sequence |
| $Del_{tr}$ | Days | Delay of activation of NPI | Fixed | 2 days |

### Output layer and data

We fit our model to time series data of reported numbers of infections $I_{S,M}$, deaths $D_M$ and occupation of ICU beds $C_M$ representing the output layer of our IO-NLDS model. Data source of infections and deaths were official reports of the Robert-Koch-Institute (RKI) in between 4th of March, 2020 and 29st of March, 2021. Number of critical cases were retrieved from the German Interdisciplinary Association of Intensive and Emergency Medicine (Deutsche Interdisziplinäre Vereinigung für Intensiv- und

Notfallmedizin e.V. – DIVI) in between 25th March, 2020 and 29st March, 2021. Time points in proximity to Christmas and the turn of the year (19th December, 2020 to 19th January, 2021) were heavily biased and therefore omitted during parameter fitting.

However, also for the considered time intervals several sources of bias need to be considered. We handled these issues as explained in the following:

*Infected cases:* We first smoothed reported numbers of infections with a sliding window of seven days centered around the time point of interest to control for strong weekly periodicity. We assume that these numbers correspond to a certain percentage $P_{S,M}$ of symptomatic patients. This is justified by the fact that the majority of reported infected cases develop symptoms (about 85% according to the RKI [9]), but there is also a large amount of asymptomatic cases (approximately 55-85% of infections (Buitrago-Garcia et al. 2020; Oran und Topol 2020; Byambasuren et al. 2020)). In the present model, we assume $P_{S,M}$ as constant. The exact formula linking states of the SECIR model with the measured numbers of infected subjects $I_{S,M}$ can be found in **Supplemental information S3** formula **S3.1**. We further accounted for delays in the reporting of case numbers by introducing a log-normally distributed delay time as explained in **Supplemental information S3**.

*Critical cases:* The number of critical COVID-19 cases (DIVI reported ICU) is available since end of March 2020 [10]. We assumed that these data are complete since 16th April, 2020 when reporting became mandatory by law in Germany. Earlier data was up-scaled from the number of reporting hospitals to the number of ICU-beds of all hospitals according to the reported ICU capacity available for 2018. We coupled the data of the critical sub-compartments $C_i$ ($i$=1,2,3) to these numbers directly.

*Deaths:* Deaths are reported by the RKI but daily reports do not reflect true death dates, which needs to be accounted for. Available daily death data of the RKI are retrospectively updated with a delay between true death date and reported date (death report delay - DRD). We assume that DRD is normally distributed with an average of 7.14 days and a standard deviation of 4 days as reported by Delgado et al. [11]. Details can be found in **Supplemental information S3**.

*Occurrence of B.1.1.7 variant:* In January 2021, the variant B.1.1.7 became endemic in Germany and quickly replaced all other variants. Onset of this development was modeled by an instantaneous influx of 5% of newly infected subjects into the $E^{Mu}$ compartment at the 26th January, 2021 estimated from published data [12].

## Parametrization

We carefully searched the literature to establish ranges for our model parameters. These ranges are used as prior constraints during parametrization of our model (**Table 2**). Parameters are then derived by fitting the predictions of the model to the reported data of infected subjects, ICU occupation and deaths using the link functions of model and data explained in the previous section. This is achieved via likelihood optimization. Likelihood is constructed using the same principles as reported [13]. In short, the likelihood consists of three major parts, namely the likelihood of deviations from prior values, the likelihood of residual deviations from the data and a penalty term to ensure that model parameters are prescribed ranges, as explained in **Supplemental information S4.** We follow a full-information approach intended to use all data collected during the epidemic as explained in **Supplemental information S5-S7**. As a result, our model fits well to the complete dynamics of the epidemic in Germany in the above mentioned time period (**Figure 2** and **Figure 3**).

To ensure identifiability of parameters, we checked a number of parsimony assumptions. For example, we assumed that the dynamical infection intensities of asymptomatic ($b_1 \cdot r_1(t)$) and symptomatic subjects ($b_2 \cdot r_2(t)$) are proportional with factor $rb_{1,2}$. We also determined Bayesian Information criteria (BIC) for different partitioning numbers of the external jump functions ($N_{crit}$ and $N_{death}$) to keep these as small as possible. Details can be found in **Supplemental information S2**.

Likelihood optimization is achieved using a stochastic approximation of an estimation-maximization algorithm (SAEM) [14]. The algortihm is based on a stochastic integration of marginal probabilities without using likelihood approximations such as linearization or quadrature approximation or sigma-point filtering [15].

Confidence intervals of model predictions are derived by Markov-Chain-Monte-Carlo simulations, i.e. alternative parameter settings were sampled from the parameter space around the optimal solution (**Supplemental information S6, S7**). We use these parameter sets to simulate alternative epidemic dynamics. This resulted in a distribution of model predictions from which empirical confidence intervals are derived.

### Implementation

The model and respective parameter estimations are implemented in the statistical software package R from which external publicly available functions are called. The model's equation solver is implemented as C++ routine and called from R code using the Rcpp package.

The code for simulation of the output layers using the reported parameter settings will be made available via our Leipzig Health Atlas : https://www.health-atlas.de/models/40 and GitHub https://github.com/GenStatLeipzig/LeipzigIMISE-SECIR [16].

## Results

### Explanation of epidemiologic dynamics

We used the full data set to explain the course of infections, ICU occupations and deaths between 4th March, 2020 and 29st March, 2021 in Germany. We identified nine fixed and 19 empirically identified time points of NPI / behavioural changes (**Supplemental Table S1**). Regarding $p_{crit}$ and $p_{death}$, we identified 18 respectively 19 time steps (See **Supplemental Tables S2, S3 and S7**). Throughout the epidemic, we observed a good agreement of our model and incident (**Figure 2**) and cumulative data (**Figure 3**). Corresponding residual errors are provided for all observables (**Supplemental Table S4**).

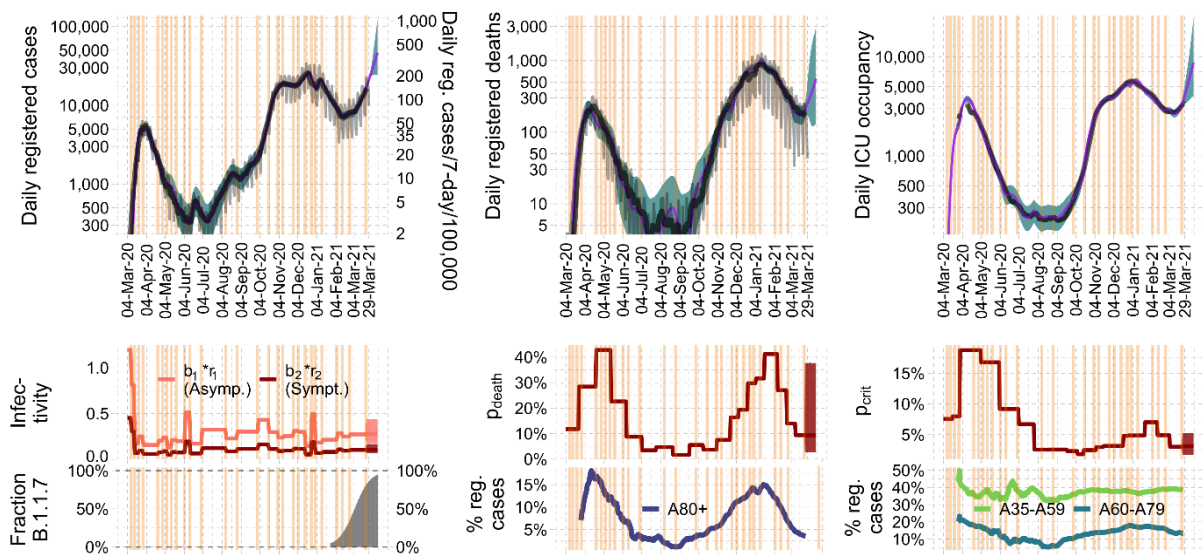

*Figure 2: Agreement of model and incident data. We show incident infections, deaths and daily ICU occupancy during the course of the epidemic in Germany in between 04th March, 2020 and 29th March, 2021. Comparison of IO-NLDS model (magenta curve) and data (thin grey curves = raw data, solid black curve = data averaged by sliding window) is provided in the upper column. A good agreement is observed (shaded area = prediction uncertainty, vertical lines = changes in NPI / contact behaviour). The middle row represents the corresponding input layer, i.e. the estimated time course of relevant input parameters such as infectivity and probabilities of critical disease course and death. Time steps correspond to the lines of changing NPI / contact behaviour as displayed in the upper row. In the lower row, we present percentages of B.1.1.7 among infected subjects (first figure), subjects older than 80 years among infected corresponding to high death tolls (second) and subjects in the age categories 35-59, respectively 60-79 among critical cases (last figure of last row).*

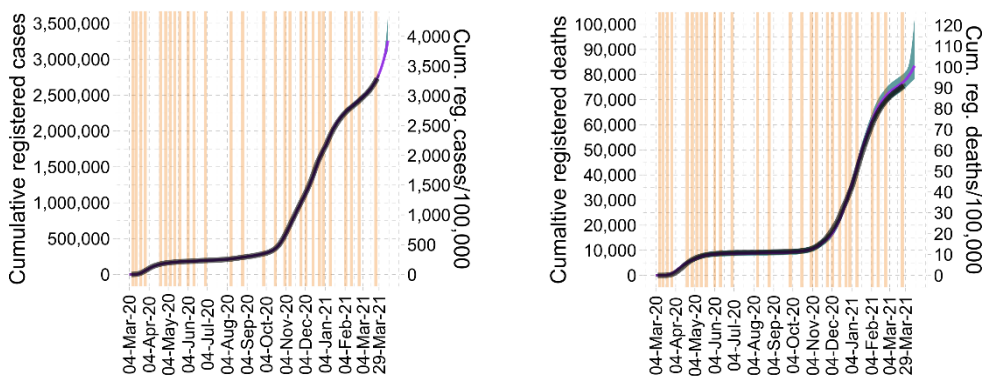

*Figure 3: Agreement of model and cumulative data. We show cumulative infections and deaths during the course of the epidemic in Germany in between 04th March, 2020 and 29th March, 2021. Comparison of IO-NLDS model (magenta curve) and data (solid black curve) is provided. A good agreement is observed (shaded area = prediction uncertainty, vertical lines = changes in NPIs / contact behaviour).*

### Parameter estimates and identifiability

Parameter estimates of the SECIR model are presented in **Table 2**, while those required to define the input layer are presented in **Table 3** and **Supplemental Tables S1, S3**. For those parameters for which we used prior information for fitting purposes, we compared the respective expected posteriors with the best priors (see **Figure 4**). Statistics are provided in **Supplemental Table S5**. No significant deviations between expected values of posteriors and priors were detected. All relative errors of parameters of the SECIR model are smaller than 10% demonstrating excellent identifiability of all epidemiologic parameters. As expected, identifiability of the external control functions is reduced. Largest standard errors of steps are in the order of 70% still demonstrating reasonable identifiability (**Supplemental Table S3**).

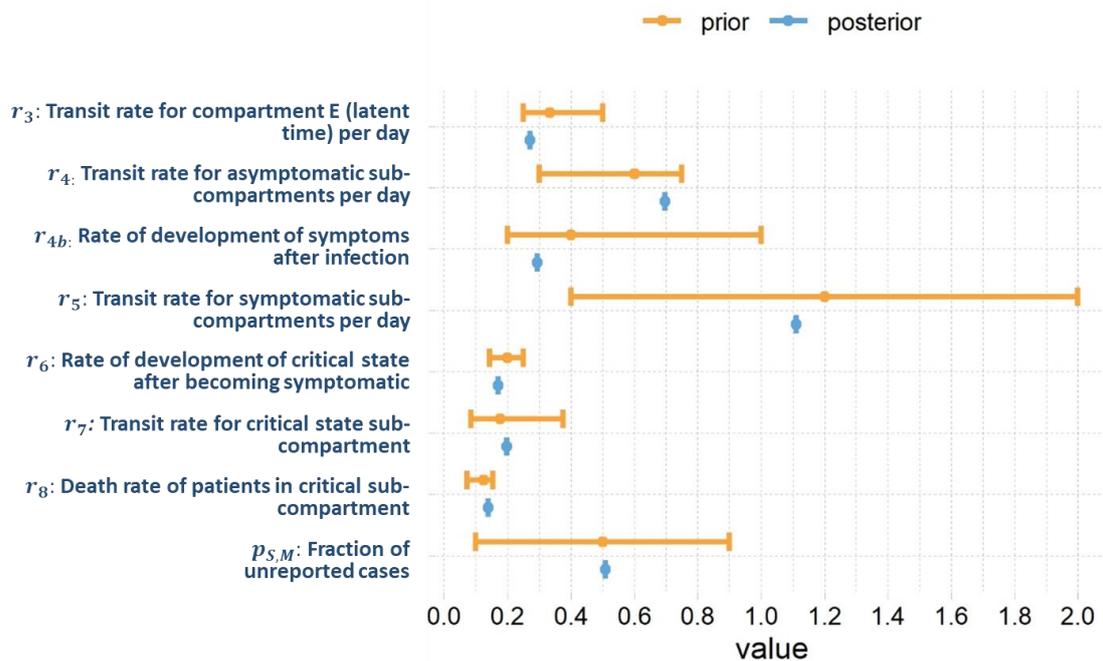

*Figure 4: Comparison of prior and posterior values of estimated parameters of the SECIR model. We present the prior distribution vs. the posterior distribution of estimated parameters of the SECIR model. Ranges for priors represent assumed minimum and maximum values. Ranges for posteriors represent 95%-confidence intervals. Numbers are provided in **Supplemental Table S8***

## Plausibilization of estimated step functions of infectivity

We estimated the infectivity as an empirical step function through the course of the epidemic. This step function should also roughly reflect NPI effectivity. We therefore compared our infectivity step function with the Governmental stringency index of NPI as estimated on the basis of Hale et al. [17]. Results are displayed at **Figure 5** and revealed a reasonable agreement.

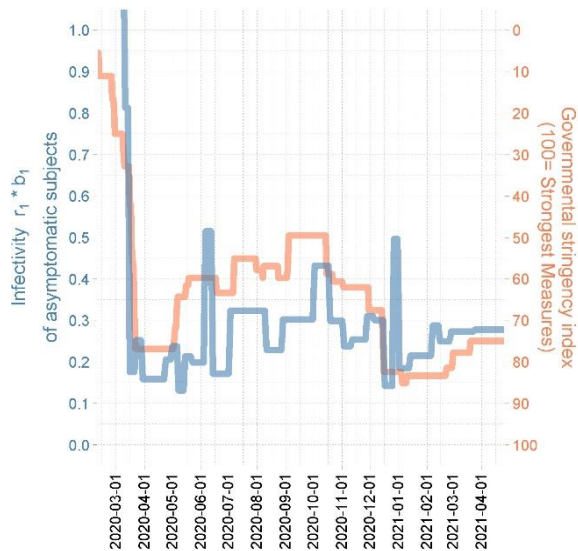

*Figure 5: Relationship between estimated step function of infectivity of asymptomatic subjects and the Federal Government stringency index (GSI).* The GSI [17] is a composite measure based on nine response indicators including school closures, workplace closures, and travel bans, rescaled to a value from 0 to 100 (100 = strictest). If policies vary at the level of federal states,, the index is shown for the state with the strictest measures. For background info see also [18].

## Model predictions

We regularly used our model to make predictions regarding the future course of the epidemic. Predictions were specifically made for the Free State of Saxony, a federal state of Germany and were published at the Leipzig Health Atlas [16]. We here present comparisons of our predictions with the actual course of the epidemic for two scenarios to demonstrate utility of our approach. Parameter values for Saxony were obtained in same way as for Germany restricting available data of infected subjects, ICU cases and deaths to this state. Estimated parameter values are presented in **Supplemental Tables S6-S8**.

While Saxony was almost spared from the first wave of SARS-CoV-2 in Germany, the second wave hit the country particularly hard resulting in the highest relative death toll of all German states (1 out of 400 inhabitants of Saxony died from COVID-19 during the second and the immediately following B.1.1.7-driven third wave). The second wave was on its peak in the middle of December 2020. A hard lock-down was initiated at this time including closure of schools, prohibition of all team-based leisure activities and night-time curfew. We were asked by the government to estimate the length of lock-down required to break the second wave. Stringency of lock-down was comparable to the first wave. Thus, we simulated four scenarios: An optimistic assumption of a lock-down efficacy of 60% reduction in infectivity, a more realistic scenario with 40% reduction, a pessimistic assumption of only 20% lock-down efficacy and, finally, 0% reduction (no lock-down) as control scenario. Results are shown in **Figure 6** and revealed a good agreement of our prediction with the actual course for the 40% scenario considered likely.

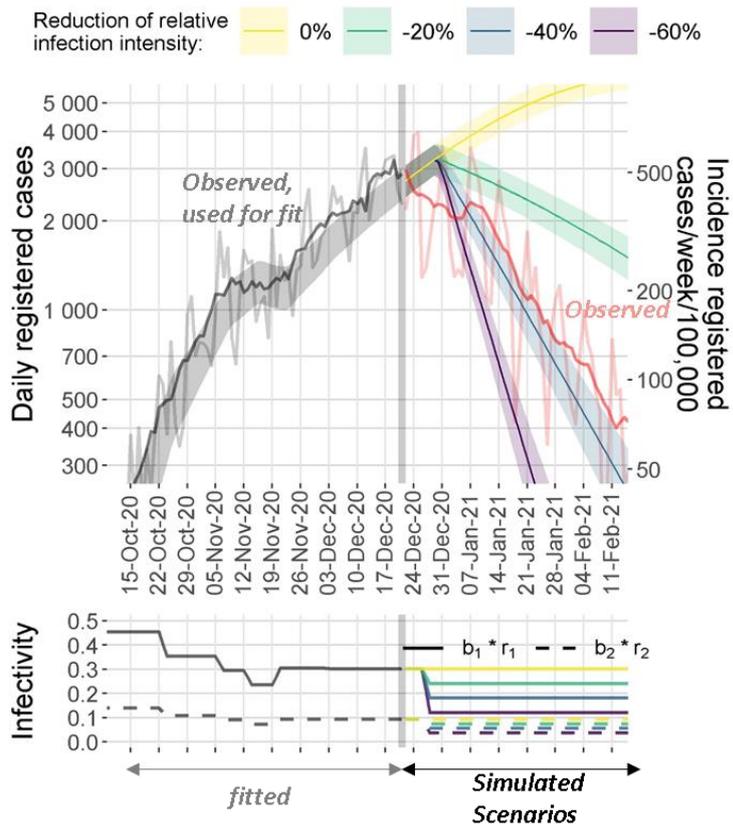

*Figure 6: Comparison of predicted and observed decline of the second wave in Saxony according to initiated lock-down.* Our model was used to fit the observed data until 21$^{st}$ of December, 2020 (shown as grey curve (raw data) and black curve (smoothed) of reported test-positives). Estimated step functions b1 and b2 describing the infectivity of asymptomatic and symptomatic subjects were reduced by 0% (yellow: no lock-down = control scenario), 20% (green: pessimistic scenario), 40% (blue: realistic scenario), and 60% (magenta: optimistic scenario) to simulate four scenarios of the future course of the epidemic under lock-down conditions. The observed numbers of test-positives after the 21$^{st}$ December, 2020 are shown in red (light red = raw data, dark red = smoothed) closely followed the expected scenario of 40% lock-down efficacy. Shaded areas represent 95% prediction intervals. The predictions and parameters were reported in our regular bulletin deposited at Leipzig Health Atlas, ID: 85AH9JMUFM-4

At the beginning of February 2021, the second wave was broken in Saxony and first relaxations of NPIs were conducted. At this time, the more virulent B.1.1.7 variant became endemic in Germany. At February 14$^{th}$, the true percentage of the B.1.1.7 variant was unknown due to lack of sequencing capacities. Moreover, there were uncertainties with respect to the increase in infectivity by the B.1.1.7 variant. We therefore simulated three scenarios (optimistic: 10% initial proportion of B.1.1.7, infectivity increased by factor 1.7, expected: 20% initial proportion, 1.8-times increase in infectivity, pessimistic: 30% initial proportion, 2-times increase in infectivity). Results are shown in **Figure 7**. The actual course of the epidemic was close to the pessimistic scenario, i.e. the second wave was directly followed by a third wave due to the B.1.1.7 variant. Indeed, later data revealed that the proportion of B.1.1.7 was already close to 30% (pessimistic assumption) at the time the simulation was performed. Moreover, our model correctly predicted the variant replacement by B.1.1.7.

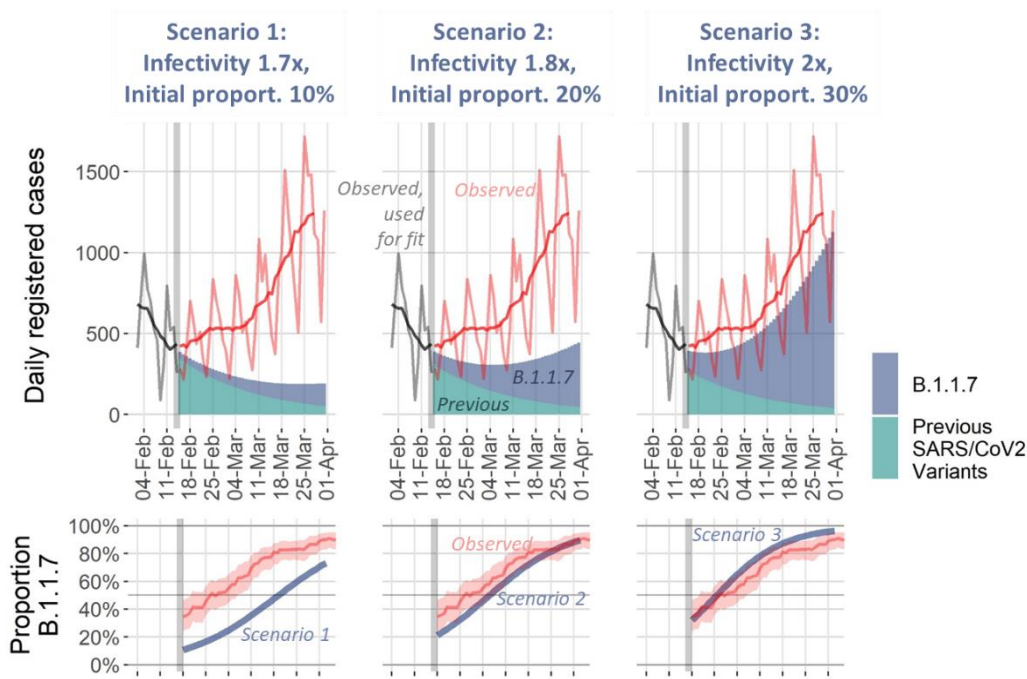

*Figure 7: Simulation of third wave scenarios for Saxony/Germany.:* <u>Upper row:</u> *The model was used to fit all observed data until 14*[th] *of February, 2021 (grey curve = raw data of reported testpositives, black curve = smoothed). Three scenarios were simulated differing in assumed initial proportion of B.1.1.7 which was not exactly known at this time point (10%, 20%, and 30%, respectively) and in the assumptions regarding increased virulence of B.1.1.7 (parameter mur = 1.7, 1.8 and 2, respectively). Predicted course of subjects infected with the respective variants are shown as shaded areas. The observed total numbers of testpositives (light red = raw data of reported testpositives, dark red = smoothed curve) closely followed the pessimistic scenario 3.* <u>Lower row:</u> *When comparing the proportion of B.1.1.7 as retrieved from [19] from 18*[th] *of July, 2021, initial proportion of B.1.1.7 was indeed close to that assumed for scenario 3. Blue curves represent 95% confidence intervals of the B.1.1.7 proportion predicted for the different scenarios. All predictions were reported in our bulletin at the 20*[th] *February, 2021 deposited at our Leipzig Health Atlas [20].*

## Discussion

In this paper, we propose a method of parametrization of COVID-19 epidemiologic models and applied it to an extended SECIR-type model to explain the course of the epidemic in Germany and one of its federal state, the Free State of Saxony. Moreover, we demonstrated how the model can be used to make relevant predictions, which could be validated on the basis of subsequent observational data.

A key idea of our approach are the embedding of differential equations based epidemic modelling into an input-output dynamical system (IO-NLDs). This has two major advantages. First, the approach allows combining explicit mechanistic models of epidemic spread and phenomenological considerations of external impacts on model parameters via the input layer. This allows parametrizing models of different complexity. For example, in our model we non-explicitly considered the effect of age structures of the diseased population by time dependent input parameters such as probabilities of critical disease courses and deaths. This could easily be replaced for example by age-structured models. We believe that such a combined empirical / mechanistic approach is well suitable for explaining the high complexity of the COVID-19 epidemic for which it is impossible to consider all relevant mechanisms explicitly and in parallel.

The second major advantage of our approach is that we assumed a non-direct link between state parameters of the embedded SECIR model and observables. This allows interposing a data model considering the known biases of the available data resources. We aimed at identifying relevant bias sources as far as possible and considered them in our proposed data models. However, these data models could be subjected to changes in the future for example if better data of COVID-19 related death will be released. Improved data models could be easily integrated into our framework.

Note that the IO-NLDS implementation translates the embedded differential equations model to a discrete scale (i.e. days in our case), which however appears to be sufficient for describing an epidemic.

There is a large plethora of epidemiologic models available for COVID-19 pandemic and the SECIR model used here is by far neither unique nor the most comprehensive one. For example, a SECIR model was also proposed by Khailaie et al [21]. Barbarossa et al. proposed a SEIR type model [22]. The Robert-Koch institute developed a model for the purpose of estimating the effect of different vaccination strategies which could easily be included into our SECIR-type models [23,24]. Although integration of differential equations-based models into our IO-NLDS context is more straightforward, this approach is also applicable to agent-based models, another often applied modelling class in the context of COVID-19 epidemic modelling [25–27] In general, the aspect of parameter estimation of such models in view of the highly biased data resources is underdeveloped and to our knowledge, no generic concept was proposed so far.

Based on our IO-NLDS formulation and data models, we parametrized our model on the basis of data of infection numbers, critical cases and deaths available for Germany and Saxony. Here, we chose a full-information approach considering all data in between start of the epidemic 4th of March, 2020 to 29$^{st}$ of March, 2021. We also applied a Bayesian learning process by considering other studies to inform model parameter's settings. Thus, we combine mechanistic model assumptions with results from other studies and observational data. This approach is very popular in pharmacology [28] and we find it important to transfer this concept to epidemiology as well. It resulted in a complex likelihood function, which is optimized on the basis of Markov-Chain Monte Carlo (MCMC) algorithms. If the likelihood has a global maximum, most of the samples typically accumulate in its vicinity after a certain number of "burn-in" steps. This allows an effective MCMC search of the best parameters estimates as well as approximation of their standard errors (standard deviations of the sample) and the degree of overfitting. However, if parameters are interdependent, MCMC algorithm samples manifolds of alternative solutions, resulting in very large standard errors of the overfitted parameters. We successfully addressed this issue by a modified version of Maire's algorithm [29]. We also applied rigorous information criteria to limit the number of steps of the input functions. As a consequence, it was possible to identify both, the fixed parameters of the SECIR model and the time-variable input functions representing changing NPI / contact behavior and age-structures.

Model parametrization resulted in a good and unbiased fit of data for the entire period for Germany. Fixed parameter values of the SECIR model did not significantly deviated from their prior values if available. It required 18 respectively 19 steps of changes of the probabilities to develop critical stage and to die respectively. A total of 13 intensification and 15 relaxation events were necessary to describe the epidemic dynamics over the time course of observation. Estimated infectivity roughly correlated with the Governmental Stringency Index [17]. We regularly contributed forecasts of our model to the German forecast Hub [27].

We also demonstrated utility of our model by several mid-term simulations of scenarios of epidemic development in Saxony, a federal state of Germany. We could show that predictions of reported infections were in the range of later observations for scenarios considered likely.

As future extensions and improvements of our model, we will consider stochastic effects on a daily scale, for example to model random influxes of cases or to model random extinctions of infection chains. These effects are relevant to be considered in times of low incidence numbers such as those observed in Germany in the summers 2020 and 2021. Our IO-NLDS framework is well suited to implement such extensions [7].

We will also include in future versions age-structure into our SECIR model, and implement a vaccination model. In the current version of the model, we assumed a constant proportion of symptomatic patients reported as infected. This does not consider for example changing testing policies (i.e. symptomatic vs. prophylactic testing). We plan to refine our model in this regard in the future. Finally, we will consider the Delta variant emerging in June 2021 [19] in the next update of our SECIR model.

In summary, the primary focus of the paper is an adequate parametrization of epidemiological models on the basis of complex, possibly biased data, as well as its coupling with structurally unknown dynamical external influences. This approach allows for a clear separation of mechanistic model compartments from random or time-dependent non-mechanistic influences and biases in the data. We believe that this approach is useful not only for the parametrization of the SECIR model presented here but also for other epidemiologic models including other disease contexts and data structures.

Acknowledgement: This project was funded in the framework of the project SaxoCOV (Saxonian COVID-19 Research Consortium). SaxoCOV was co-financed with tax funds on the basis of the budget passed by the Saxon state parliament. Presentation of data, model results and simulations was funded by the NFDI4Health Task Force COVID-19 (www.nfdi4health.de/task-force-covid-19-2) within the framework of a DFG-project (LO-342/17-1).

## References

1. Adiga A, Dubhashi D, Lewis B, Marathe M, Venkatramanan S, Vullikanti A. Mathematical Models for COVID-19 Pandemic: A Comparative Analysis. J Indian Inst Sci. 2020:1–15. Epub 2020/10/30. doi: 10.1007/s41745-020-00200-6 PMID: 33144763.
2. Flaxman S, Mishra S, Gandy A, Unwin HJT, Mellan TA, Coupland H, et al. Estimating the effects of non-pharmaceutical interventions on COVID-19 in Europe. Nature. 2020; 584:257–61. Epub 2020/06/08. doi: 10.1038/s41586-020-2405-7 PMID: 32512579.
3. Bo Y, Guo C, Lin C, Zeng Y, Li HB, Zhang Y, et al. Effectiveness of non-pharmaceutical interventions on COVID-19 transmission in 190 countries from 23 January to 13 April 2020. Int J Infect Dis. 2021; 102:247–53. Epub 2020/10/29. doi: 10.1016/j.ijid.2020.10.066 PMID: 33129965.
4. Harris JE. Overcoming Reporting Delays Is Critical to Timely Epidemic Monitoring: The Case of COVID-19 in New York City. ; 2020.
5. Böttcher S, Oh D-Y, Staat D, Stern D, Albrecht S, Wilrich N, et al. Erfassung der SARS-CoV-2-Testzahlen in Deutschland (Stand 2.12.2020). Robert Koch-Institut; 2020.
6. McCulloh I, Kiernan K, Kent T. Improved Estimation of Daily COVID-19 Rate from Incomplete Data. 2020 Fourth International Conference on Multimedia Computing, Networking and Applications (MCNA). IEEE; 19.10.2020 - 22.10.2020. pp. 153–8.
7. Georgatzis K, Williams CKI, Hawthorne C. Input-Output Non-Linear Dynamical Systems applied to Physiological Condition Monitoring. Proceedings of the 1st Machine Learning for Healthcare Conference 2016: PMLR; 19 August 2016 through 20 August 2016 [updated 2016 Aug 19 through 2016 Aug 20].
8. Wise J. Covid-19: New coronavirus variant is identified in UK. BMJ. 2020; 371:m4857. Epub 2020/12/16. doi: 10.1136/bmj.m4857 PMID: 33328153.
9. COVID-19. clinical aspects. Robert Koch Institute [updated 16 Sep 2021; cited 16 Sep 2021]. Available from:

https://www.rki.de/DE/Content/InfAZ/N/Neuartiges_Coronavirus/Daten/Klinische_Aspekte.html.
10. Tagesreport-Archiv. Deutsche Interdisziplinäre Vereinigung für Intensiv- und Notfallmedizin (DIVI) e.V. [updated 29 Mar 2021; cited 29 Mar 2021]. Available from: https://www.divi.de/divi-intensivregister-tagesreport-archiv.
11. George Delgado, MD, FAAFP, John Safranek, MD, PhD, Bill Goyette, BS, MSEE, Richard Spady P. Reported versus Actual Date of Death. "Reported" versus "Actual": Two Different Things. Available from: https://covidplanningtools.com/reported-versus-actual-date-of-death/.
12. Bericht zu Virusvarianten von SARS-CoV-2 in Deutschland, insbesondere zur Variant of Concern (VOC) B.1.1.7. Robert Koch Institute [updated 3 Mar 2021; cited 3 Mar 2021]. Available from: https://www.rki.de/DE/Content/InfAZ/N/Neuartiges_Coronavirus/DESH/Bericht_VOC_2021-03-03.pdf?__blob=publicationFile.
13. Kheifetz Y, Scholz M. Modeling individual time courses of thrombopoiesis during multi-cyclic chemotherapy. PLoS Comput Biol. 2019; 15:e1006775. Epub 2019/03/06. doi: 10.1371/journal.pcbi.1006775 PMID: 30840616.
14. Kuhn E, Lavielle M. Coupling a stochastic approximation version of EM with an MCMC procedure. ESAIM: PS. 2004; 8:115–31. doi: 10.1051/ps:2004007.
15. Konstantinos Georgatzis, Christopher K.I. Williams, Christopher Hawthorne. Input-Output Non-Linear Dynamical Systems applied to PhysiologicalCondition Monitoring. ; 2016.
16. Meineke FA, Löbe M, Stäubert S. Introducing Technical Aspects of Research Data Management in the Leipzig Health Atlas. Stud Health Technol Inform. 2018; 247:426–30.
17. Hale T, Angrist N, Goldszmidt R, Kira B, Petherick A, Phillips T, et al. A global panel database of pandemic policies (Oxford COVID-19 Government Response Tracker). Nat Hum Behav. 2021; 5:529–38. Epub 2021/03/08. doi: 10.1038/s41562-021-01079-8 PMID: 33686204.
18. COVID-19 Government Response Tracker [updated 1 Aug 2021; cited 1 Aug 2021]. Available from: https://www.bsg.ox.ac.uk/research/research-projects/covid-19-government-response-tracker.
19. Mullen JL, Tsueng G, Latif AA, Alkuzweny M, Cano M, Haag E, et al. Outbreak.info. a standardized, open-source database of COVID-19 resources and epidemiology data [cited 18 Jul 2021]. Available from: https://outbreak.info.
20. Aktuelle Entwicklung der COVID-19 Epidemie in Leipzig und Sachsen. Bulletin 14 vom 20.02.2021. Available from: https://www.imise.uni-leipzig.de/sites/www.imise.uni-leipzig.de/files/files/uploads/Medien/bulletin_n14_covid19_sachsen__2021_02_22_v11.pdf.
21. Khailaie S, Mitra T, Bandyophadhyay A, Schips M, Mascheroni P, Vanella P, et al. Development of the reproduction number from coronavirus SARS-CoV-2 case data in Germany and implications for political measures. ; 2020.
22. Barbarossa MV, Fuhrmann J, Meinke JH, Krieg S, Varma HV, Castelletti N, et al. Modeling the spread of COVID-19 in Germany: Early assessment and possible scenarios. PLoS One. 2020; 15:e0238559. Epub 2020/09/04. doi: 10.1371/journal.pone.0238559 PMID: 32886696.
23. Der Heiden M an, Buchholz U. Modellierung von Beispielszenarien der SARS-CoV-2-Epidemie 2020 in Deutschland. Robert Koch-Institut; 2020.
24. Scholz S, Waize M, Weidemann F, Treskova-Schwarzbach M, Haas L, Harder T, et al. Einfluss von Impfungen und Kontaktreduktionen auf die dritte Welle der SARS-CoV-2-Pandemie und perspektivische Rückkehr zu prä-pandemischem Kontaktverhalten. 2021. doi: 10.25646/8256.
25. Kucharski AJ, Klepac P, Conlan AJK, Kissler SM, Tang ML, Fry H, et al. Effectiveness of isolation, testing, contact tracing, and physical distancing on reducing transmission of SARS-CoV-2 in different settings: a mathematical modelling study. Lancet Infect Dis. 2020; 20:1151–60. Epub 2020/06/16. doi: 10.1016/S1473-3099(20)30457-6 PMID: 32559451.

26. Quilty BJ, Clifford S, Hellewell J, Russell TW, Kucharski AJ, Flasche S, et al. Quarantine and testing strategies in contact tracing for SARS-CoV-2: a modelling study. Lancet Public Health. 2021; 6:e175-e183. Epub 2021/01/21. doi: 10.1016/S2468-2667(20)30308-X PMID: 33484644.
27. Bracher J, Wolffram D, Deuschel J, Görgen K, Ketterer JL, Ullrich A, et al. A pre-registered short-term forecasting study of COVID-19 in Germany and Poland during the second wave. Nat Commun. 2021; 12:5173. Epub 2021/08/27. doi: 10.1038/s41467-021-25207-0 PMID: 34453047.
28. Friberg LE, Henningsson A, Maas H, Nguyen L, Karlsson MO. Model of chemotherapy-induced myelosuppression with parameter consistency across drugs. J Clin Oncol. 2002; 20:4713–21. doi: 10.1200/JCO.2002.02.140 PMID: 12488418.
29. Maire F, Friel N, Mira A, Raftery AE. Adaptive Incremental Mixture Markov Chain Monte Carlo. J Comput Graph Stat. 2019; 28:790–805. Epub 2019/06/07. doi: 10.1080/10618600.2019.1598872 PMID: 32410811.

# Supporting Information

## S1 Equations of the SECIR model

We here present the equations of the SECIR model serving as hidden layer of our Input-Output non-linear dynamical system (IO-NLDS). To fit in this context, the ordinary differential equations of the SECIR model are approximated by a difference equation system describing compartment changes at single days, i.e. time-steps $\Delta t$ equals one day. Compartments of the model are explained in **Table 1** of the main paper. Parameters are explained in **Table 2** of the main paper.

$$\frac{\Delta Sc}{\Delta t} = -influx - Influx_E - Influx_E^{Mu}$$

$$\frac{\Delta E}{\Delta t} = influx + Influx_E - r_3 \cdot E$$

$$\frac{\Delta I_{A,1}}{\Delta t} = r_3 \cdot E - \left(X(r_{4,b}, r_4, p_{symp}) \cdot r_{4,b} + \left(1 - X(r_{4,b}, r_4, p_{symp})\right) \cdot r_4\right) \cdot I_{A,1}$$

$$\frac{\Delta I_{A,2}}{\Delta t} = \left(1 - X(r_{4,b}, r_4, p_{symp})\right) \cdot r_4 \cdot I_{A,1} - r_4 \cdot I_{A,2}$$

$$\frac{\Delta I_{A,3}}{\Delta t} = r_4 \cdot I_{A,2} - r_4 \cdot I_{A,3}$$

$$\frac{\Delta I_{S,1}}{\Delta t} = X(r_{4,b}, r_{4,b}, p_{symp}) \cdot r_{4,b} \cdot I_{A,1} - \left(X(r_6, r_5, p_{crit}) \cdot r_6 + \left(1 - X(r_6, r_5, p_{crit})\right) \cdot r_5\right) \cdot I_{S,1}$$

$$\frac{\Delta I_{S,2}}{\Delta t} = \left(1 - X(r_6, r_5, p_{crit})\right) \cdot r_5 \cdot I_{S,1} - r_5 \cdot \left(1 - p_{death,S}\right) \cdot I_{S,2} - p_{death,S} \cdot r_8 \cdot I_{S,2}$$

$$\frac{\Delta I_{S,3}}{\Delta t} = \left(1 - p_{death,S}\right) \cdot r_5 \cdot I_{S,2} - r_5 \cdot I_{S,3}$$

$$\frac{\Delta C_1}{\Delta t} = X(r_6, r_5, p_{crit}) \cdot r_6 \cdot \left(I_{S,1} + I_{S,1}^{Mu}\right) - \left(X(r_8, r_7, p_{death}) \cdot r_8 + \left(1 - X(r_8, r_7, p_{death})\right) \cdot r_7\right) \cdot C_1$$

$$\frac{\Delta C_2}{\Delta t} = \left(1 - X(r_8, r_7, p_{death})\right) \cdot r_7 \cdot C_1 - r_7 \cdot C_2$$

$$\frac{\Delta C_3}{\Delta t} = r_7 \cdot C_2 - r_7 \cdot C_3$$

$$\frac{\Delta R}{\Delta t} = r_5 \cdot \left(I_{S,3} + I_{S,3}^{Mu}\right) + r_7 \cdot C_3 + r_4 \cdot \left(I_{A,3} + I_{A,3}^{Mu}\right)$$

$$\frac{\Delta D}{\Delta t} = X(r_8, r_7, p_{death}) \cdot r_8 \cdot C_1 + p_{death,S} \cdot r_8 \cdot \left(I_{S,2} + I_{S,2}^{Mu}\right)$$

(S1.1)

With the abbreviations

$$Influx_E = r_1 \cdot b_1 \cdot Sc \cdot \left(I_{A,1} + I_{A,2} + I_{A,3}\right) + r_2 \cdot b_2 \cdot Sc \cdot \left(I_{S,1} + I_{S,2} + I_{S,3}\right)$$

$$Influx_E^{Mu} = Mur \left(r_1 \cdot b_1 \cdot Sc \cdot \left(I_{A,1}^{Mu} + I_{A,2}^{Mu} + I_{A,3}^{Mu}\right) + r_2 \cdot b_2 \cdot Sc \cdot \left(I_{S,1}^{Mu} + I_{S,2}^{Mu} + I_{S,3}^{Mu}\right)\right).$$

At this, we assume that asymptomatic and symptomatic compartments have different time dependent infectivity ($r_1 * b_1$, $r_2 * b_2$), with *b1* and *b2* later defined on the basis of time-dependent NPI / behavioural changes and, for parsimony, $b_1 = b_2$. Hence, the ratio of the products ($r_1 * b_1$) and ($r_2 * b_2$) is assumed constant.

The functions *X* represent decisions regarding the further disease course defined below.

Analogously, the equations for the concurrent variant compartments are as follows:

$$\frac{\Delta E^{Mu}}{\Delta t} = Influx_E^{Mu} - r_3 \cdot E^{Mu}$$

$$\frac{\Delta I_{A,1}^{Mu}}{\Delta t} = r_3 \cdot E^{Mu} - \left(X(r_{4,b}, r_4, p_{symp}) \cdot r_{4,b} + \left(1 - X(r_{4,b}, r_4, p_{symp})\right) \cdot r_4\right) \cdot I_{A,1}^{Mu}$$

$$\frac{\Delta I_{A,2}^{Mu}}{\Delta t} = \left(1 - X(r_{4,b}, r_4, p_{symp})\right) \cdot r_4 \cdot I_{A,1}^{Mu} - r_4 \cdot I_{A,2}^{Mu}$$

$$\frac{\Delta I_{A,3}^{Mu}}{\Delta t} = r_4 \cdot I_{A,2}^{Mu} - r_4 \cdot I_{A,3}^{Mu}$$

$$\frac{\Delta I_{S,1}^{Mu}}{\Delta t} = X(r_{4,b}, r_{4,b}, p_{symp}) \cdot r_{4,b} \cdot I_{A,1}^{Mu} - \left(X(r_6, r_5, p_{crit}) \cdot r_6 + \left(1 - X(r_6, r_5, p_{crit})\right) \cdot r_5\right) \cdot I_{S,1}^{Mu}$$

$$\frac{\Delta I_{S,2}^{Mu}}{\Delta t} = \left(1 - X(r_6, r_5, p_{crit})\right) \cdot r_5 \cdot I_{S,1}^{Mu} - r_5 \cdot (1 - p_{death,S}) \cdot I_{S,2}^{Mu} - p_{death,S} \cdot r_8 \cdot I_{S,2}^{Mu}$$

$$\frac{\Delta I_{S,3}^{Mu}}{\Delta t} = (1 - p_{death,S}) \cdot r_5 \cdot I_{S,2}^{Mu} - r_5 \cdot I_{S,3}^{Mu}$$

(S1.2)

In some of the compartments, there is a decision between recovery or deterioration of the disease course. For example patients in $C_1$ can either die or recover with different rates $r_i$ and $r_j$. To model this decision, we split the number of patients in the compartment by a respective probability $p$ and determine the decision factor $X$ as follows.

$$\frac{r_i X}{r_j (1-X)} = \frac{p}{1-p},$$

Thus

$$X(r_i, r_j, p) = \frac{r_j p}{r_i \cdot (1-p) + r_j \cdot p} \tag{S1.3}$$

To start the epidemic in Germany, we assumed an entry of infected cases by a linearly decreasing function starting at 4th of March, 2020 and becoming zero at 10th of March, 2020. Occurrence of the B.1.1.7 variant was initialized by a single influx to $E^{Mu}, I_{A,1}^{Mu}, I_{A,2}^{Mu}, I_{A,3}^{Mu}, I_{S,1}^{Mu}, I_{S,2}^{Mu}, I_{S,3}^{Mu}$ at the 26th of February, 2021. We assume that the ratio between these influxes are the same as those for the normal variant compartments at this day. The sum of $E^{Mu}, I_{A,1}^{Mu}, I_{A,2}^{Mu}, I_{A,3}^{Mu}, I_{S,1}^{Mu}, I_{S,2}^{Mu}, I_{S,3}^{Mu}$ at 26th of February, 2021 was fitted to be 5.3% of the corresponding sum of the normal compartments at the same day.

## S2 Input layer

The input layer of our IO-NLDS is designed to describe the effects of non-pharmaceutical interventions (NPI) and other impacts on infectivity such as behavioral changes, changes in age-structure, testing policy, seasonal effects or larger outbreaks (abbreviated as NPI / contact behaviour). Since these effects typically affect contact matrices in different ways, we model this phenomenologically by time dependent reductions or increases of infection rates caused by symptomatic and asymptomatic subjects as explained in S.1.

We make the following assumptions for non-pharmaceutical interventions:

1. We introduce the relative infectivity function $b(t)$, which changes according to NPI / contact behaviour modifications. This is modelled by a linear increase (in case of relaxation) or decrease (in case of tightening) within a fixed time $Del_{tr}$ of two days. Otherwise, $b(t)$ is constant. We denote $\{T_{tr,s}\}_{s=1}^{N_{tr}}$ as the time points with changes in non-pharmaceutical interventions with $N_{tr}$ the total number of time points with changes. We collected dates of changing non-pharmaceutical intervention measures for Germany based on Government decisions, changing testing policies as well as events with impact on epidemiological dynamics such as holidays and sudden outbreaks(such as thin peak of new infections in June affected mostly workers of the meat industry). We also assumed additional time points with changes determined by BIC.

Again, for the sake of parsimony, we assume that the relative infection intensities of asymptomatic ($b_1(t)$) and symptomatic subjects ($b_2(t)$) are the same, hence, respective proportionality $rb_{1,2}$ is constant and estimated during model fitting.

Thus, $b(t)$ is defined as follows:

$$b(t) = \begin{cases} b_{tr,s-1}, & t \in [T_{tr,s-1} + Del_{tr}, T_{tr,s}] \\ \frac{b_{tr,s-1}}{Del_{tr}} \cdot (Del_{tr} - t + T_{tr,s}) + \frac{b_{tr,s}}{Del_{tr}} \cdot (t - T_{tr,s}), & t \in [T_{tr,s}, T_{tr,s} + Del_{tr}] \\ b_{tr,s}, & t \in [T_{tr,s} + Del_{tr}, T_{tr,s+1}] \\ b_{tr,0} = 1 \end{cases}, \quad (S2.1)$$

$$b_1(t) = b(t)$$
$$b_2(t) = b_1(t)$$

The time point $t=0$ corresponds to the 3$^{rd}$ of March, 2020.

2. Likewise, rates towards critical disease states and deaths are also assumed to vary through the course of the epidemic due to changes in testing policy resulting in different percentages of unreported cases and asymptomatic subjects, changes in age distribution of infected subjects, improvement of patient care due to new treatment options and due to possible over-stretched medical resources (not the case in Germany but other countries). In our model, this is also accounted for phenomenologically by assuming the probabilities $p_{crit}$ and $p_{death}$ as time-dependent input parameters.

We assume that both functions are step functions:

$$p_{crit}(t) = p_{crit,0} \cdot \sum_{j=0}^{N_{crit}-1} \alpha_{crit,j} \cdot \chi_{[T_{pcrit,j}, T_{pcrit,j+1})}(t)$$
(S2.2)

where $\{T_{pcrit,j}\}_{j=1}^{N_{crit}}$ are empirical dates and $\{\alpha_{crit,j}\}_{j=1}^{N_{crit}}$ the respective relative changes of $p_{crit}$. Both, $\{T_{pcrit,j}\}_{j=1}^{N_{crit}}$ as well as $\{\alpha_{crit,j}\}_{j=1}^{N_{crit}}$ are parameters to be estimated. The initial value of $p_{crit}$ is $p_{crit,0}$. Functions $\chi_{[t_j,t_{j+1})}(t)$ are indicator functions being 1 in the interval $[T_{pcrit,j}, T_{pcrit,j+1})$ and 0 else.

The step functions for $p_{death}(t)$ and $p_{death,S}(t)$ are defined analogously:

$$p_{death}(t) = p_{death,0} \cdot \sum_{j=0}^{N_{death}-1} \alpha_{death,j} \cdot \chi_{[T_{pdeath,j}, T_{pdeath,j+1})}(t)$$
$$p_{death,S}(t) = p_{death,S,0} \cdot p_{death}(t)$$
, (S2.3)

The partitions of $p_{crit}$ and $p_{death}$ are assumed independent. Respective numbers of jumps $N_{crit}$ and $N_{death}$ can differ.

In order to find an optimal tradeoff between parsimony and goodness of fit we calculated Bayesian Information criteria (BIC) for different partition numbers $N_{tr}$, $N_{crit}$ and $N_{death}$ and chose partitions minimizing BIC.

When new data become available, we attempt to update the numbers of partitions every two weeks by considering adding a new break-point within the last month. The time point as well as the corresponding jump value are considered as two new parameters. We added a new break point only if it improves BIC after the updated parameters estimation.

## S3 Output layer

We here describe, how the state parameters of the hidden SECIR model are linked with data via the output layer of the IO-NLDS.

*Modeling of daily registered infected cases $I_{S,M}$:* The total number of daily registered infected cases $I_{S,M}$ is coupled to the efflux of the first asymptomatic compartments $I_{A,1}$ and $I_{A,1}^{Mu}$ towards symptomatic compartments multiplied with $P_{S,M}$.

$$I_{S,M}(T) = \sum_{t=0}^{T} P_{S,M} \cdot X(r_{4,b}, r_4, p_{symp}) \cdot r_{4,b} \cdot \left(I_{A,1}(t) + I_{A,1}^{Mu}(t)\right)(t). \tag{S.3.1}$$

However, we assume a delay of the registered vs. reported cases by introducing an empirical distribution of the reporting delay: When a person has a positive PCR SARS-CoV-2 test result at a date $d_1$, this will result in a registered case at a later date $d_2$. The difference $d=d_2-d_1$ is the reporting delay and is assumed log-normally distributed. This distribution of delays is determined on the basis of data provided by the Robert-Koch-Institute from the period 27[th] of April, 2020 to 13[th] of November, 2020. The parameters of this distribution are derived by minimizing the Kullback-Leibler divergence between the parametric representation and the empirical distribution. Results are displayed in supplemental figure S1.

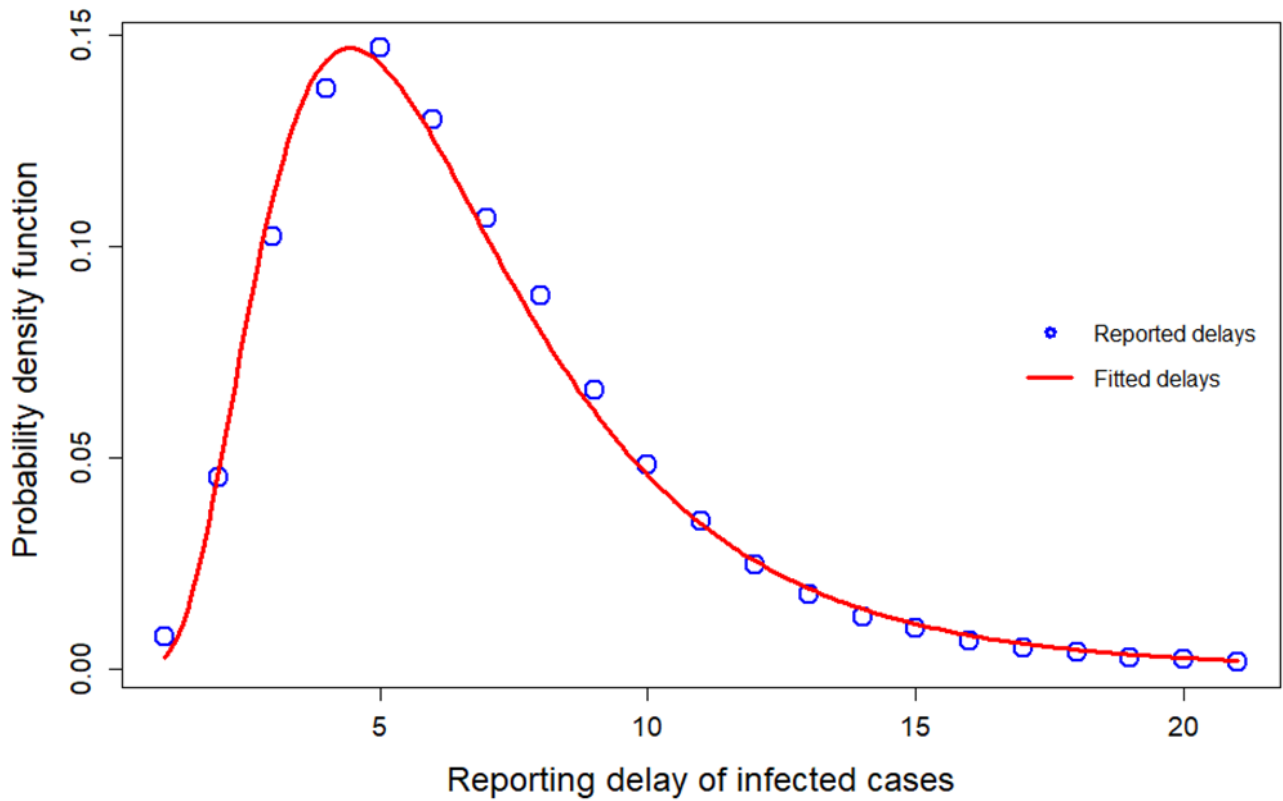

*Supplemental figure S1: Approximation of reporting delay by a log-normal distribution:* We present the log-normal distribution best fitting the empirical distribution of reporting delays. Estimated parameters of the log-normal distribution are as follows: µ=1.77 days, σ=0.531 days.

<u>Modeling delay of death reporting:</u> In contrast to the newly infected cases, neither information of delays in COVID-19 associated death reporting nor actual dates of deaths were available to us. Therefore, we used a data model proposed by Delagdo et al. [11]. In detail, we assumed that the delay is normally distributed with an average of 7.14 days and a standard deviation of 4 days.

Since we consider time as an integer, we discretize this normal distribution by the approximation DRD(d)= $\frac{N_{7.14,4}(d)}{\sum_{i=1}^{100} N_{7.14,4}(i)}$ for integers $d \leq 100$ and 0 else, where N is the Gaussian distribution function with mean 7.14 days and standard deviation of 4 days, i.e. we neglect delays larger than 100 days. Using this approximation, we derive the actual number of new deaths at time point *t*:

$$D_a(t) = \sum_{t1<t} Dr(t) \cdot DRD(t - t1).$$

Here, $D_r(t)$ is the number of reported new deaths at time *t*. The function $D_a(t)$ is linked to our compartment *D*.

## S4 Parameter estimation

Free parameters of the model are determined by minimizing the negative log-likelihood function of observed data. The likelihood is constructed in analogy to [13]. It constitutes of the sum of three components:

$$nLL = nLL^{pri} + nLL^{resid} + Constr. \tag{S4.1}$$

The terms $nll_i^{pri}$ and $nll_i^{resid}$ correspond to prior constrains of parameters and to the residual errors of the data as explained below in detail. The term *Constr* is a penalty term to keep values in eligible ranges or orders (see "Penalization"). We assume independence between parameters throughout.

Parameter distributions and transformations: Most of the parameters are confined to certain ranges. During estimation (with possible prior constrains), we transform these parameters to the space of real numbers. We assume that these transformed values are normally distributed during Markov-Chain Monte Carlo (MCMC) sampling (see below). To ensure this, parameters confined to a finite interval (a,b) are transformed by the *logit*-function. Parameters with positive values are transformed by a log-normal transformation. Thus,

$$\varphi_s = h_k(\psi_s)$$
$$h_s(\psi_s) = \begin{cases} e^{\psi_s}, & \text{for parameters} > 0 \\ a + (b-a) \cdot \frac{e^{\psi_s}}{1+e^{\psi_s}}, & \text{for parameters within } [a,b] \end{cases}, \quad s = 1, \dots, N_{par}, \tag{S4.2}$$

where $\varphi_s$ is the *s*-th parameter and $\psi_s$ is the respective transformed parameter and $N_{par}$ is the total number of parameters to be estimated.

The likelihood contribution of the priors $nll_i^{pri}$ is defined as follows:

$$nLL^{pri} = \sum_{s=1}^{N_{par}} \delta_s \cdot \frac{\left(\psi_s - \psi_s^{pri}\right)^2}{\omega_{pri,s}^2}. \tag{S4.3}$$

where $\delta_s$ equals 1, if a prior is assumed for the *s*-th parameter and 0 otherwise. The prior information is represented by the "best value" $\psi_s^{pri}$ and an uncertainty expressed as standard deviation of possible values $\omega_{pri,s}$. We assume that parameter estimates are random variables normally distributed around their respective prior values. Thus,

$$\psi_s \sim N(\psi_s^{pri}, \omega_{pri,s}) = N\left(h_s(\varphi_s^{pri})^{-1}, \omega_{pri,s}\right). \tag{S4.4}$$

Prior "best values" and ranges of parameters are provided at **Table 2** of the main paper. The uncertainties $\omega_{pri,k}$ are set to 2 for all parameters. This heuristic setting is based on a tradeoff between avoidance of overfitting including implausible parameter values and good data fitting properties.

Penalization: We penalize with a high value of $10^8$ in cases when times of non-pharmaceutical interventions are either too close (closer than 3 days) or non-monotonic. In the same way, we penalize too high dynamical $p_{death}$ values (more than 0.66) by multiplication of max($p_{death}$-0.66,0) with 100.

Residual errors of observed vs. predicted data: We fit data for daily registered cases, cumulative registered cases, deaths, cumulative deaths and ICU occupation as explained in sub-section "output layer" and the methods section of the main paper. The respective term of the negative log-likelihood $nll_i^{resid}$ corresponds to the residual errors of these data. Thus,

$$nLL^{resid} = \sum_{Y_{out}} \left( we_{dY_{out}} \cdot \sum_{j=1}^{N_{dY_{out}}} \frac{\left(dY_{out,S}^{tr_{dY_{out}}}(t_{j,dYout},\psi) - dY_{out,D}^{tr_{dY_{out}}}(t_{j,dYout})\right)^2}{a_{dY_{out}}^2} + we_{cumul} \cdot we_{Y_{out}} \cdot \right.$$

$$\left. \sum_{j=1}^{N_{dY_{out}}} \frac{\left(Y_{out,S}^{tr_{Y_{out}}}(t_{j,dYout},\psi) - Y_{out,D}^{tr_{Y_{out}}}(t_{j,Yout})\right)^2}{a_{Y_{out}}^2} \right), \tag{S4.5}$$

where $Y_{out}$ represents the output layers ("dY" corresponds to daily counts, while "Y" corresponds to cumulative counts). Subscript *S* denotes simulation results, while *D* corresponds to the data. We sum the likelihoods of the three outputs considered (infected subjects, critical cases and deaths). Thus, $Y_{out}$ represents one of these three entities *x* with number of data points $N_x$ at time points $\{t_{j,x}\}$ (j=1,…,N$_x$) and residual errors $a_x$. We introduce weights $we_{cumul}$ for the cumulative terms as compared with the daily counts and set it to 0.2. The cumulative terms were introduced to avoid biases of cumulative data occurring after fitting daily data only. Cumulative data for ICU occupation were not fitted, i.e. $we_{ICU} = 0$. The parameter $tr_{Y_{out}}$ corresponds to the power transformation used to compare model and data. In the present model version, it is set to 0.5. It constitutes a tradeoff between fitting precision of large and small numbers. All weights $we_{dY_{out}}$ and $we_{Y_{out}}$ were set to 1.

Thus, we assume that that for each output and for each data point the entities $dY_{out,D}^{tr_{dY_{out}}}$ and $Y_{out,D}^{tr_{dY_{out}}}$ are normally distributed random variables around respective simulated values with standard deviations being the respective residual errors:

$$\begin{aligned}dY_{out,D}^{tr_{dY_{out}}}(t_{j,dYout}) &\sim N\left(dY_{out,S}^{tr_{dY_{out}}}(t_{j,dYout},\psi), a_{dY_{out}}\right) \\ Y_{out,D}^{tr_{dY_{out}}}(t_{j,Yout}) &\sim N\left(Y_{out,S}^{tr_{Y_{out}}}(t_{j,Yout},\psi), a_{Y_{out}}\right)\end{aligned}, \tag{S4.6}$$

The algorithm to minimize the negative log-likelihood is explained in the next section. Differences of estimated values and their respective priors can be tested by calculating Z-scores $\frac{\psi_s - \psi_s^{pri}}{\omega_{pri,s}}$.

## S5. Algorithm for parameter estimations and prediction sampling

Due to nonlinearity of (Formulas S4.1, S4.4, S4.5), parameters $\psi$ and residual errors $\theta$ cannot be estimated simultaneously. For such situations an expectation-maximization (EM) algorithm was proposed by Dempster et al. [30]. This algorithm is a widely applied approach for the iterative computation of maximum likelihood estimates in incomplete-data statistical problems. In detail, the random parameters $\psi = \{\psi_s\}_{s=1}^{N_{par}}$ are considered as non-observed data, while observed data $y$ in our case are defined as follows:

$$y = \left\{\{I_M^D(t_{j,IM})\}_{j=1}^{N_{I_M}}, \{D^D(t_{j,D})\}_{j=1}^{N_D}, \{dI_M^D(t_{j,IM})\}_{j=1}^{N_{dI_M}}, \{dD^D(t_{j,D})\}_{j=1}^{N_{dD}}, \{dICU_M^D(t_{j,ICU})\}_{j=1}^{N_{dICU}}\right\},$$
(S5.1)

Complete data of the model is $(y, \psi)$. The unknown residual errors $\theta$ describe the uncertainty of parameters $\psi$.

Therefore $nLL$ is a marginal likelihood. The complete likelihood nLL is defined as follows:

$$nLL(y; \theta) = \int_\Omega nLL(y, \psi; \theta) d\psi.$$
(S5.2)

The EM algorithm minimizes $nLL(y; \theta)$ iteratively: At the $k$-th iteration of EM, the expectation step computes the conditional expectation of the complete log-likelihood $Q_k(\theta) = E(nLL(y, \psi; \theta)|y, \theta_{k-1})$ by generating $\psi^{(k)}$ based on previous estimates $\theta_{k-1}$, and the maximization step computes the value $\theta_k$ maximizing $Q_k(\theta)$. The EM sequence $(\theta_k)$ converges to a stationary point under general regularity conditions [30].

In nonlinear cases, the expectation step cannot be performed in a closed form. Therefore, we applied the Stochastic Approximation algorithm of EM (SAEM). SAEM is a maximum likelihood estimator of the population parameters [14] based on stochastic integration of marginal probabilities without likelihood approximation such as linearization or quadrature approximation or sigma-point filtering [15]. Our implementation is inspired by and is very similar to that of earlier versions of Monolix (Lixoft) software http://lixoft.com/.

The stochastic approximation version of standard EM algorithm (SAEM) proposed by [14] replaces the usual E-step at an iteration $k$ by a stochastic procedure as follows:

1. Simulation step: draw $m_k$ realizations of $\psi^{(k)} = \{\psi_s^{(k)}\}_{s=1}^{N_{par}}$ from the conditional distribution $p(\cdot|y; \theta_k)$ using MCMC algorithm.
2. Stochastic approximation: update $Q_k(\theta)$

$$Q_k(\theta) = Q_{k-1}(\theta) + \gamma_k \cdot \left(\frac{1}{m_k}\sum_{j=1}^{m_k} log\left(p(y, \psi^{(k)}; \theta)\right) - Q_{k-1}(\theta)\right)$$
(S5.3)

   where $\gamma_k$ is a decreasing sequence of positive numbers.
3. Maximization-step: update $\theta_k$ according to

$$\theta_{k+1} = Arg\,\max_\theta(Q_k(\theta)).$$
(S5.4)

Remarks:

1. Our stochastic approximation step is an improved version of the stochastic approximation of the integration of marginal distribution on the multidimensional domain $\Omega$ of possible parameter values:

$$Q_k(\theta) = E\left(log(p(y, \psi; \theta))|y, \theta_{k-1}\right) = \int_\Omega log\left(p(y, \psi^{(k)}; \theta_{k-1})\right) d\psi^{(k)}$$
(S5.5)

2. In analogy to Monolix software, we selected $\gamma_k$ as follows:

$$\gamma_k = 1, \quad k \le K_1$$
$$\gamma_k = \frac{1}{k-K_1+1}, \quad k > K_1. \tag{S5.6}$$

We choose $K_1$ equal to 4 and run the algorithm until convergence with a tolerance 0.1% of estimates of population parameters (see below).

3. We performed MCMC sampling 4000 times at each stage with a burn-in phase of 1000 steps. Thus, $m_k = 3000$.

Exact estimates of different components of $\theta_k$ are:

$$a_{dIM}^{(k)} = \sqrt{\frac{1}{N_{dIM}} \sum_{j=1}^{N_{dIM}} \int_\Omega \left( dI_M^S(t_{j,IM}, \psi^{(k)}) - dI_M^D(t_{j,IM}) \right)^2 d\psi^{(k)}}$$

$$a_{dD}^{(k)} = \sqrt{\frac{1}{N_{dD}} \sum_{j=1}^{N_{dD}} \int_\Omega \left( dD^S(t_{j,D}, \psi^{(k)}) - dD^D(t_{j,D}) \right)^2 d\psi^{(k)}} \tag{S5.7}$$

$$a_{dICU}^{(k)} = \sqrt{\frac{1}{N_{dICU}} \sum_{j=1}^{N_{dICU}} \int_\Omega \left( dICU^S(t_{j,ICU}, \psi^{(k)}) - dICU^D(t_{j,ICU}) \right)^2 d\psi^{(k)}}$$

Therefore, respective stochastic approximations and maximizations $\theta_k$ are as follows:

$$s_{1,j,k} = s_{1,j,k-1} + \gamma_k \cdot \left( \frac{1}{m_k} \sum_{r=1}^{m_k} \left( dI_M^S(t_{j,IM}, \psi^{(k,r)}) - dI_M^D(t_{j,IM}) \right)^2 - s_{1,j,k-1} \right), \quad j = 1, \cdots, N_{dIM}$$

$$a_{dIM}^{(k)} = \sqrt{\frac{\sum_{j=1}^{N_{dIM}} s_{1,i,j,k}}{N_{dIM}}}$$

(S5.8)

$$s_{2,j,k} = s_{2,j,k-1} + \gamma_k \cdot \left( \frac{1}{m_k} \sum_{r=1}^{m_k} \left( dD^S(t_{j,IM}, \psi^{(k,r)}) - dD^D(t_{j,IM}) \right)^2 - s_{2,j,k-1} \right), \quad j = 1, \cdots, N_{dD}$$

$$a_{dD}^{(k)} = \sqrt{\frac{\sum_{j=1}^{N_{dD}} s_{2,i,j,k}}{N_{dD}}}$$

(S5.9)

$$s_{3,j,k} = s_{3,j,k-1} + \gamma_k \cdot \left( \frac{1}{m_k} \sum_{r=1}^{m_k} \left( dICU^S(t_{j,IM}, \psi^{(k,r)}) - dICU^D(t_{j,IM}) \right)^2 - s_{3,j,k-1} \right), \quad j = 1, \cdots, N_{dICU}$$

$$a_{dICU}^{(k)} = \sqrt{\frac{\sum_{j=1}^{N_{dICU}} s_{3,i,j,k}}{N_{dICU}}}$$

(S5.10)

In the same way, the respective terms for the cumulative data approximations are derived.

## S6 MCMC algorithm for the expectation step

Markov chain Monte Carlo (MCMC) methods comprise a class of algorithms for sampling from a probability distribution [31]. By constructing a Markov chain that has the desired distribution as its equilibrium distribution, one can obtain a sample of the desired distribution by recording states from the chain. It is well known that a proper choice of a proposal distribution for MCMC methods is a crucial factor for convergence of the algorithm [32]. For the sake of increasing the acceptance rate, a number of adaptive Metropolis (AM) algorithms were proposed by different groups. Here the proposal distribution is learned along the process using the full information cumulated so far. We implemented the adaptive MCMC version with Gaussian proposal distribution described in (Haario et al. 2001) as well as adaptive incremental Mixture MCMC [29] called AIMM, which we modified slightly. Strictly speaking, these methods are not really Markov chains, because proposal distribution of the next step depends on all preceding states $\{\vec{X}_t\}_0^t$ rather than only the previous one. The algorithm of Haario et al. is simpler and it assumes the existence of a global minimum of nLL. In contrast, the algorithm of Maire et al. could be useful for cases when the nLL has a complex topology due to overfitting.

Let $\pi$ denote the target distribution (i.e. likelihood) given by formula (S4.1). At each iteration step the new parameter vector $Y$ is generated by a transition kernel representing the proposal distribution. This candidate vector is accepted with probability

$$\alpha(\vec{X}_{t-1}, Y) = \min\left(1, \frac{\pi(Y)}{\pi(\vec{X}_{t-1})}\right). \tag{S6.1}$$

The transition kernel of Haario's MCMC version is an empirical covariance matrix of previous samples stabilized by an identity matrix multiplied by a small number $\varepsilon$:

$$C_t^{(k)} = s_d \cdot \left(cov(\psi^{(k,1)}, \cdots, \psi^{(k,t-1)}) + \varepsilon \cdot Id\right), \quad t \leq m_k, \tag{S6.2}$$

where $t$ is a sampling number and $s_d = \frac{2.4}{\sqrt{N_{par}^{ind}}}$. Here, we choose $\varepsilon = 0.0001$. This parameter is required to ensure ergodic property of the Markov chain. When $k>1$, we added samples from the previous step to the covariance matrix:

$$C_t^{(k)} = s_d \cdot \left(cov(\psi^{(k-1,1)}, \cdots, \psi^{(k-1,m_k)}, \psi^{(k,1)}, \cdots, \psi^{(k,t-1)}) + \varepsilon \cdot Id\right), \quad t \leq m_k. \tag{S6.3}$$

At each iteration $k$ we used a sample from the previous iteration providing a small value of nLL as starting point.

In the AIMM, the proposal distributions $Q_t$ are mixtures of multivariate normal distributions. Roughly spoken, this is a generalization of Haario's algorithm when multiple local minima of the nLL exist in few clusters. The candidate vector is accepted with probability

$$\alpha(\vec{X}_{t-1}, Y) = \min\left(1, \frac{\pi(Y)/Q_t}{\pi(\vec{X}_{t-1})/Q_{t-1}}\right), \tag{S6.4}$$

This algorithm minimizes discrepancies between proposal and target probability i.e. a sequence $\{Q_t\}$ converges to $\pi$ by approximating it through mixtures of multivariate normal distributions. The elements of this series $Q_t$ are defined as follows:

$$Q_t = \frac{\sum_{l=1}^{M_t} \beta_l \cdot \varphi_l}{\sum_{l=1}^{M_t} \beta_l},$$

where $M_t$ is the number of components at the iteration $t$. The elements $\{\varphi_1, \cdots, \varphi_{M_t}\}$ represent the incremental mixture components, $\{\beta_1, \cdots, \beta_{M_t}\}$ are the respective weights. Each mixture component consists of a mean vector and a covariance matrix. The sampling from $Q_t$ proceeds as follows: We choose the $r$-th component with a probability $\frac{\beta_r}{\sum_{l=1}^{M_t} \beta_l}$ by generating a uniformly distributed random number and accepting the $r$-th component if this number is in between $\left(\frac{\beta_{r-1}}{\sum_{l=1}^{M_t} \beta_l}, \frac{\beta_r}{\sum_{l=1}^{M_t} \beta_l}\right]$ if $r>1$ or in between $\left[0, \frac{\beta_r}{\sum_{l=1}^{M_t} \beta_l}\right]$ if $r>1$. After the choice of the $r$-th component, a random parameter vector $Y$ is generated around the $r$-th mean according to the $r$-th covariance matrix as in Haario's algorithm. If $Y$ is accepted, it becomes $\vec{X}_t$. $\vec{X}_t$ can either stay in the $r$-th cluster or give origin for the new cluster $\varphi_{M_t}$ with $M_t = 1 + M_{t-1}$. A new cluster is created when the match of $\varphi_{M_t}$ to the $r$-th cluster is insufficient based on Mahalanobis distance. If $\vec{X}_t$ stays in the *r*-th cluster, it updates the *r*-th covariance matrix in a similar way as in Haario's algorithm [32], formula (S6.3).

We here modified the conditions for new cluster formation compared to [29] as follows. In our algorithm, a new cluster is formed when one of the following conditions hold:

- The Mahalanobis distance of $\vec{X}_t$ to the cluster from which it was generated is less than 0.025 or larger than 0.975. That is $\vec{X}_t$ diverges significantly from the current multivariate normal distribution of the *r*-th cluster
- $\pi(\vec{X}_t)$ is significantly larger than π of the current cluster center. That is $\vec{X}_t$ does not correspond to the local maximum of π in the neighbourhood of the *r*-th cluster.

If one of the above conditions holds, $\vec{X}_t$ becomes the center of a new cluster. The respective Gaussian component is the covariance matrix of the *r*-th (i.e. previous) cluster. This matrix will be further updated every time when new members of the new cluster are accepted in future proposals.

The weights $\{\beta_1, \cdots, \beta_{M_t}\}$ are proportional to π of the respective cluster centers to the power of γ, where γ is a positive number less than 1. All weights are updated every time when a new cluster emerges.

In summary, AIMM accepts proposals with discrepancies to the target distribution. As a consequence, proposal distributions are multivariate normal mixtures. Every cluster's mean is a local maximum of π. Sampling of proposal distributions from clusters depend on π. New clusters emerge when an accepted proposal either significantly diverges from the respective cluster's probability or when a significantly better optimal value is found in this cluster.

After thorough comparison of adaptive MCMC and the adaptive incremental mixture MCMC, we found the latter to be superior. It finds higher values of π in a shorter time. It also generates higher acceptance rates (0.2-0.3 versus 0.1) and finds more alternative solutions. We therefore used this method for our parameter estimations.

## S7 MCMC simulation for prediction and controlling goodness of fit

The estimates of residual errors are determined at the last step and are used for MCMC sampling of parameters $\psi$. The resulting means and standard deviations are considered as respective average estimates and their standard errors. Simulations of these parameter samples provides a set of alternative predictions. From these, we collected the best fitting solution, the average solution and confidence intervals for different confidence limits α.

## S8 Justification of prior parameters and ranges

We here provide justifications of assumed prior values and parameter ranges. Details of parameter fitting can be found in **supplement material section S2**.

Initial influx of people per day *Influx*: The initial influx was estimated from the data without prior assumptions to a value of 6,937 people per day in order to initialize the simulation. Later, the parameter is no longer relevant for simulation outcomes.

Infection rate through asymptomatic subjects per day $r_1$: This infection rate was estimated from the data without prior assumptions. It represents the basic transmission probability of the SARS-CoV-2 virus from an asymptomatic infectious person to a susceptible contact.

Infection rate through symptomatic subjects per day $r_2$: This infection rate was estimated from the data without prior assumptions. It represents the basic transmission probability of the SARS-CoV-2 virus from a symptomatic infectious person to a susceptible contact.

Relative infection intensity of asymptomatic subjects per day $b_1(t)$: The infectivity of asymptomatic infected subjects was assumed as a time-dependent step-function due to changing NPIs / contact behaviour and other factors influencing infection probabilities. Steps were estimated from the data without prior assumptions.

Relative infection intensity of symptomatic subjects per day $b_2(t)$: The infectivity of symptomatic infected subjects was assumed as a time-dependent step-function due to changing NPIs / contact behaviour and other factors influencing infection probabilities. Steps were estimated from the data without prior assumptions.

Ratio of $b_1(t)$ and $b_2(t)$ ($rb_{1,2}$). We assumed a fixed ratio of the infectivities of asymptomatic and symptomatic infected subjects for the sake of parsimony. The ratio was estimated from the data without prior assumptions.

Fixed dates for updates of infectivity functions: We used several fixed dates of changes in infectivity functions due to known changes in NPIs, testing policy or outbreaks. Note that even fixed time points were checked for necessity to assume changes in infectivity for the sake of parsimony, i.e. respective steps were only assumed if significantly improving model fit. The first three fixed time-points, $tr_1$ (March 10th, 2020), $tr_2$ (March 15th, 2020), and $tr_3$ (March 22nd, 2020) reflect German governmental interventions including regulation of the size of public events, travel restrictions, and contact restriction. Fixed time-points $tr_6$ (April 30th, 2020), $tr_7$ (May 7th, 2020), and $tr_8$ (May 21st, 2020) reflect German governmental interventions related to the step-wise relaxation of NPIs, in particular regarding leisure sports, contacts, and schools. Time point $tr_{17}$ (November 2nd, 2020) reflects governmental NPIs in response to the German second wave, including restrictions of public life and social contacts, also referred as "soft lockdown". Time point $tr_{21}$ (December 16th, 2020) reflects further stricter governmental NPIs in response to the ongoing increase of the German second wave, also referred as "hard lockdown", strongly limiting public and private contacts including school closures. Finally, time point $tr_{28}$ (February 23rd, 2021) reflects release of many governmental NPIs in response to the decline of the German second wave.

Transit rate for compartment E (latent time) $r_3$: The transition rate $r_3$ for the compartment of exposed subjects is the inverse of the latent time, i.e. the time being infected but not yet infectious. The mean of the prior distribution for the latent time was set to 3 days and the minimum and maximum of the distribution was set to 2 and 4 days, respectively, in accordance with previous reports [23]. Note that minimum and maximum of a parameter's distribution in this section always refer to the distribution of the mean of the parameter, not of the distribution of the parameter itself. Further justification of this parameter is discussed in the following when considering the rate $r_{4b}$.

Transit rate for asymptomatic sub-compartments $r_4$: The transition rate $r_4$ for the asymptomatic infectious compartment to the recovered compartment is a third of the inverse of the time being asymptomatic and infectious, as this compartment is split into three sub-compartments. The mean of the prior distribution of $r_4$ was set to 3/5 per day and the minimum and maximum of the distribution was set to 3/10 and 3/4 per day, respectively. These values are based on general considerations regarding timelines of the germinal center reaction [33] and further supported by reports from the literature estimating relevant infectiousness periods in general or asymptomatic / mildly symptomatic COVID-19 patients as in between 3.5 to 9.5 days [34–36].

Rate of development of symptoms after infection $r_{4b}$: The inverse of this rate is equal to the time from being infectious to start of developing symptoms. The mean of the prior distribution of $r_{4b}$ was set to 1/2.5 per day and the minimum and maximum of the distribution was set to 1/5 and 1/1 per day, respectively. This is in line with previous reports [23,37,38]. Note that the serial interval, i.e. the average time between successive cases in a chain of transmission is composed of two parameters of our model. In detail, the serial interval is the sum of the average latent time ($1/r_3$) and half of the average time being infectious when assuming random occurrence of subsequent infections during time of infectiousness. Exemplarily, if the serial interval would be considered in a scenario where symptomatic individuals are immediately and effectively quarantined, the serial interval would be $1/r_3+0.5*1/r_{4b}$. The serial interval was estimated by the RKI [39] to have a median of 4 days (interquartile range 3-5 days) based on literature [40–43], which is in accordance with our choices for $r_3$ and $r_{4b}$. However, the serial interval (and other parameters like the time being infectious) are to some extent also time dependent, reflecting e.g. behavioral changes. Although we do not model a time dependence for these specific parameters, our model can, to a certain extent, cope for this by data-driven adaptation of other time dependent parameters like $b_1$ and $b_2$.

Probability of developing symptomatic disease after infection $p_{symp}$: This probability was estimated from the literature, reporting a percentage of symptomatic COVID-19 cases in between 55% and 85% [44–46]. We used a percentage of 50% as mean of the prior distribution, accounting for the fact that minor symptoms are frequently ignored or considered as symptoms of a common cold. Minimum and maximum was set to 0.3 and 0.8, respectively.

Transit rate of symptomatic sub-compartments $r_5$: The transition rate $r_5$ for the three symptomatic sub-compartments towards recovery is a third of the inverse of time being symptomatic and infectious. The mean of the prior distribution of $r_5$ was set to 3/2.5 per day and the minimum and maximum of the distribution was set to 3/7.5 and 3/1.5 per day, respectively. These values are based on the assumption that symptomatic and asymptomatic subjects are similar with respect to time of contagiousness. Hence, values of the distribution of $r_5$ equals that of $r_4$ subtracted by the mean value of $r_{4b}$.

Rate of development of critical state after becoming symptomatic $r_6$: The inverse of this rate is assumed equal to the time of developing a critical state after being infectious and symptomatic. The mean of the prior distribution of $r_6$ was set to 1/5 per day and the minimum and maximum of the distribution was set to 1/7 and 1/4 per day, respectively, according to previous reports [23,47–49]. Note that the probability of people becoming critical is affected by the function $p_{crit}$.

Probability of becoming critical after developing symptoms $p_{crit}$: This probability is assumed as a time-dependent step-function reflecting for example changing age-distributions of infected subjects or treatment efficacy. Steps were estimated from the data within the range of 0 to 1 without assuming a specific prior. The initial value of $p_{crit,0}$ was estimated as 0.075, which is within the range of reported values [23,50].

Transit rate for critical state sub-compartments $r_7$: The transition rate $r_7$ for the critical state sub-compartments is a third of the inverse of the time treated on intensive care unit (ICU) for survivors, as the critical compartment is also split into three sub-compartments. The mean of the prior distribution of $r_7$ was set to 3/17 per day and the minimum and maximum of the distribution was set to 3/35 and 3/8 per day, respectively. These values are informed by previous reports focusing on data of 35 to 79 year old patients, the most frequent population on ICU [23,51–53].

Death rate of patients in critical sub-compartment $r_8$: This transition rate represents the rate from the first ICU sub-compartment to the death compartment. It is the inverse of the time of patients on ICU that pass away. The mean of the prior distribution of $r_8$ was set to 1/8 per day and the minimum and maximum of the distribution was set to 1/14 and 1/6.5 per day, respectively. This reflects the shorter time on ICU for patients with fatal disease outcome informed by previous reports [47,54,55]. The number of people with fatal disease course is affected by two additional parameters $p_{death}$ and $p_{death,S}$ explained below.

Probability of death after becoming critical $p_{death}$: This is the probability of death for patients at ICU. It is assumed as a time-dependent step-function estimated from data. Values are restricted within the range 0 to 1 without specific prior assumptions. Changes in time reflect for example changes in age-composition of ICU patients as well as changes in treatment regimens. The initial value is $p_{death,0}$ = 0.118.

Probability of death after developing symptoms without becoming critical $p_{death,S}$: To reflect COVID-19 related deaths outside of ICU (especially relevant for the oldest age-groups [51]), we introduced the probability $p_{death,S}$ of transitioning from the second symptomatic sub-compartment to the death compartment. This probability was estimated from the data $p_{death,S}$ = 0.0448.

Fraction of unreported cases $p_{S,M}$: For the fraction of infected cases that are symptomatic but not reported, we used a prior distribution with a mean of 0.5, a minimum of 0.3 and a maximum of 0.9. This choice was informed by studies of SARS-CoV-2 seroprevalence in Germany [56,57]. Note that the total percentage of unreported infected people is $1-p_{S,M}*p_{symp}$ according to the definition of $p_{symp}$.

Factor of increased infectivity of new virus variant $mur$: This factor is multiplied to $r_1$ and $r_2$ reflecting higher infectivity of the B.1.1.7 variant compared to the previous variants. This parameter was estimated from sequencing data reporting the dynamics of the increase of variant B.1.1.7 in the UK, Denmark, Belgium, Suisse, and the United States, available from https://github.com/tomwenseleers/newcovid_belgium/ and Germany, available from "Mutationstracking-Projekt von Sven Schmidt" at https://tinyurl.com/36xnmxat. Thereby, $mur$ was calibrated to match the observed average dynamic of the increase of B.1.1.7 across countries resulting in a value of $mur$ = 1.7.

## S.9 Supplemental tables
### Parameter values for Germany

**Supplemental Table S1: Time points of changes in infectivity and respective steps.** We used fixed (known due to Governmental decisions or random events) and estimated time points of NPI / contact behaviour changes and events and respective changes in infectivity of asymptomatic subjects. We provide estimates and relative standard errors of the infectivity. For estimated time points, we also provide the respective standard error (last column).

| Number | Type of NPI / contact behaviour change | Estimated new infectivity | Relative standard error, % | Date | Source of time point | Standard error (days) |
|---|---|---|---|---|---|---|
| 1 | Intensification | 0.676 | 0.738 | 10.03.2020 | Fixed | - |
| 2 | Intensification | 0.150 | 3.99 | 15.03.2020 | Fixed | - |
| 3 | Relaxation | 0.214 | 0.711 | 22.03.2020 | Fixed | - |
| 4 | Intensification | 0.131 | 2.79 | 29.03.2020 | Estimated | 0.280 |
| 5 | Relaxation | 0.172 | 2.78 | 23.04.2020 | Estimated | 0.164 |
| 6 | Relaxation | 0.200 | 0.462 | 30.04.2020 | Fixed | - |
| 7 | Intensification | 0.109 | 5.78 | 07.05.2020 | Fixed | - |
| 8 | Relaxation | 0.177 | 5.13 | 14.05.2020 | Fixed | - |
| 9 | Intensification | 0.163 | 0.278 | 22.05.2020 | Estimated | 0.322 |
| 10 | Relaxation | 0.434 | 0.644 | 05.06.2020 | Estimated | 0.387 |
| 11 | Intensification | 0.142 | 3.67 | 13.06.2020 | Estimated | 0.360 |
| 12 | Relaxation | 0.270 | 2.80 | 01.07.2020 | Estimated | 0.251 |
| 13 | Intensification | 0.193 | 1.06 | 11.08.2020 | Estimated | 0.244 |
| 14 | Relaxation | 0.256 | 1.01 | 28.08.2020 | Estimated | 0.264 |
| 15 | Relaxation | 0.357 | 1.06 | 01.10.2020 | Estimated | 0.119 |
| 16 | Intensification | 0.246 | 0.967 | 19.10.2020 | Estimated | 0.334 |
| 17 | Intensification | 0.198 | 3.98 | 02.11.2020 | Fixed | - |
| 18 | Relaxation | 0.213 | 0.991 | 11.11.2020 | Estimated | 1.08 |
| 19 | Relaxation | 0.256 | 0.550 | 24.11.2020 | Estimated | 0.262 |
| 20 | Intensification | 0.248 | 1.61 | 01.12.2020 | Estimated | 0.303 |
| 21 | Intensification | 0.118 | 2.18 | 16.12.2020 | Fixed | - |
| 22 | Relaxation | 0.421 | 1.40 | 26.12.2020 | Estimated | 0.238 |
| 23 | Intensification | 0.154 | 3.48 | 01.01.2021 | Estimated | 0.118 |
| 24 | Relaxation | 0.182 | 11.0 | 12.01.2021 | Estimated | - |
| 25 | Relaxation | 0.237 | 3.09 | 06.02.2021 | Estimated | - |
| 26 | Intensification | 0.211 | 4.08 | 15.02.2021 | Estimated | - |
| 27 | Relaxation | 0.233 | 2.98 | 25.02.2021 | Estimated | - |
| 28 | Relaxation | 0.228 | 23.2 | 18.03.2021 | Fixed | - |

**Supplemental Table S2: Determination of the number of time steps of input step functions.** We analyzed different numbers of steps for the step functions $p_{crit}$ ($N_{crit}$) and $p_{death}$ ($N_{death}$). Npar = number of parameters to be estimated, nLL = negative log-likelihood, BIC = Bayesian information criterion. A total of 1,714 data points were analysed (348 new cases and death cases measurements daily and cumulative, 322 measurements of daily critical cases). The combination $N_{crit}$ = 18 and $N_{death}$ = 19 resulted in the lowest BIC, i.e. best compromise between model parsimony and fit. The best solution resulted from estimation of 134 parameters as follows: 15 basic parameters **(Table S5)**, 28 of infectivity changes at 19 time points **(Table S1)**, 18 values for $\alpha_{crit,i}$ with respect to 17 time points and 19 values for $\alpha_{death,i}$ with respected to 18 time points. Alternative assumptions on $N_{crit}$ and $N_{death}$ resulted in respective changes of the total number of parameters.

| $N_{crit}$ | $N_{death}$ | $N_{par}$ | nLL | BIC |
|---|---|---|---|---|
| **18** | **19** | 134 | 2620 | **6238** |
| 17 | 17 | 132 | 2661 | 6298 |
| 17 | 19 | 133 | 2645 | 6280 |
| 18 | 18 | 133 | 2633 | 6256 |
| 19 | 20 | 136 | 2616 | 6245 |
| 19 | 19 | 135 | 2618 | 6241 |

**Supplemental Table S3: Step functions of $p_{crit}$ and $p_{death}$.** We present estimates for the single steps of the functions $p_{crit}$ and $p_{death}$ at the specified dates and respective standard errors. We also provide the standard error of the estimated time point (last column).

| Parameter | Description | Estimate | Relative standard error. % | Date respective controls | Standard error (days) |
|---|---|---|---|---|---|
| $\alpha_{crit,1}$ | Relative values of $p_{crit}$ starting at the respective date | 1.05 | 0.317 | 20.03.2020 | 0.0844 |
| $\alpha_{crit,2}$ | | 2.48 | 3.18 | 01.04.2020 | 0.14 |
| $\alpha_{crit,3}$ | | 2.24 | 3.46 | 06.05.2020 | 1.06 |
| $\alpha_{crit,4}$ | | 1.22 | 3.20 | 04.06.2020 | 2.03 |
| $\alpha_{crit,5}$ | | 0.884 | 0.626 | 06.07.2020 | 3.75 |
| $\alpha_{crit,6}$ | | 0.344 | 2.07 | 30.07.2020 | 1.14 |
| $\alpha_{crit,7}$ | | 0.340 | 0.381 | 24.08.2020 | 6.94 |
| $\alpha_{crit,8}$ | | 0.301 | 4.25 | 20.09.2020 | 0.705 |
| $\alpha_{crit,9}$ | | 0.238 | 1.15 | 06.10.2020 | 1.52 |
| $\alpha_{crit,10}$ | | 0.330 | 1.03 | 23.10.2020 | 1.42 |
| $\alpha_{crit,11}$ | | 0.382 | 0.801 | 08.11.2020 | 0.870 |
| $\alpha_{crit,12}$ | | 0.419 | 1.70 | 20.11.2020 | 6.20 |
| $\alpha_{crit,13}$ | | 0.633 | 1.53 | 23.12.2020 | 1.43 |
| $\alpha_{crit,14}$ | | 0.651 | 1.51 | 01.01.2021 | 0.506 |
| $\alpha_{crit,15}$ | | 0.929 | 1.12 | 22.01.2021 | 3.43 |
| $\alpha_{crit,16}$ | | 0.647 | 3.41 | 13.02.2021 | 3.08 |
| $\alpha_{crit,17}$ | | 0.394 | 0.972 | 05.03.2021 | 6.15 |
| $\alpha_{crit,18}$ | | 0.441 | 62.8 | 18.03.2021 | - |
| $\alpha_{death,1}$ | Relative values of $p_{death}$ starting at | 2.39 | 1.88 | 26.03.2020 | 0.164 |
| $\alpha_{death,2}$ | | 3.58 | 1.17 | 23.04.2020 | 0.283 |
| $\alpha_{death,3}$ | | 1.94 | 4.55 | 19.05.2020 | 1.45 |
| $\alpha_{death,4}$ | | 0.743 | 1.19 | 10.06.2020 | 0.393 |

| $\alpha_{death,5}$ | the respective date | 0.296 | 3.25 | 05.07.2020 | 5.72 |
| --- | --- | --- | --- | --- | --- |
| $\alpha_{death,6}$ | | 0.401 | 0.635 | 27.07.2020 | 6.15 |
| $\alpha_{death,7}$ | | 0.142 | 1.22 | 25.08.2020 | 4.51 |
| $\alpha_{death,8}$ | | 0.473 | 7.46 | 17.09.2020 | 1.20 |
| $\alpha_{death,9}$ | | 0.314 | 6.39 | 08.10.2020 | 1.56 |
| $\alpha_{death,10}$ | | 0.638 | 0.966 | 01.11.2020 | 2.18 |
| $\alpha_{death,11}$ | | 1.41 | 0.748 | 22.11.2020 | 1.17 |
| $\alpha_{death,12}$ | | 1.64 | 2.53 | 11.12.2020 | 1.89 |
| $\alpha_{death,13}$ | | 2.54 | 1.35 | 29.12.2020 | 0.499 |
| $\alpha_{death,14}$ | | 2.66 | 3.01 | 07.01.2021 | 6.02 |
| $\alpha_{death,15}$ | | 3.48 | 6.42 | 18.01.2021 | 1.43 |
| $\alpha_{death,16}$ | | 2.31 | 4.80 | 05.02.2021 | 0.794 |
| $\alpha_{death,17}$ | | 1.22 | 3.15 | 27.02.2021 | 2.315 |
| $\alpha_{death,18}$ | | 0.807 | 2.15 | 09.03.2021 | 3.75 |
| $\alpha_{death,19}$ | | 1.09 | 69.1 | 19.03.2021 | - |

**Supplemental Table S4: Residual errors of observables.** We present the residual errors of fitting our model to the time frame 3rd of March, 2020 to 21st of March, 2021. dIM = daily incident cases, IM = cumulative, dICU = daily occupation of ICU beds dD=daily deathy, D = cumulative delay. Case numbers were square root transformed, i.e. units of values are cases to the power of 0.5.

| Parameter | Value for Germany | Value for Saxony |
| --- | --- | --- |
| $a_{dIM}$ | 3.62 | 0.921 |
| $a_{IM}$ | 5.69 | 1.03 |
| $a_{dICU}$ | 1.19 | 0.442 |
| $a_{dD}$ | 3.04 | 0.377 |
| $a_D$ | 0.99 | 1.14 |

**Supplemental Table S5: Parameter estimates and comparison with average priors.** We present estimated parameters of the SECIR model and initial conditions of control parameters and their respective standard errors for Germany. We also perform a formal comparison of estimates and expected priors using t-Test.

| Parameter | Description | Posterior estimate | Relative standard error, % | Prior value | p-value |
| --- | --- | --- | --- | --- | --- |
| influx | Initial influx of infections into compartment $E$ until first interventions | 3171 | 3.12 | - | - |
| $r_1$ | Infection rate through asymptomatic subjects | 1.19 | 0.582 | - | - |
| $r_3$ | Transit rate for compartment E (latent time) | 0.272 | 0.0571 | 1/3 | 0.213 |
| $r_4$ | Transit rate for asymptomatic sub-compartments | 0.636 | 0.734 | 3/5 | 0.429 |

| | | | | | | |
|---|---|---|---|---|---|---|
| $r_{4,b}$ | Rate of development of symptoms after infection | 0.456 | 2.17 | 1/2.5 | 0.346 | |
| $r_5$ | Transit rate for symptomatic sub-compartments | 0.946 | 2.33 | 3/2.5 | 0.499 | |
| $r_6$ | Rate of development of critical state after being symptomatic | 0.186 | 0.405 | 1/5 | 0.457 | |
| $r_7$ | Transit rate for critical state sub-compartment | 0.159 | 0.336 | 3/17 | 0.402 | |
| $r_8$ | Death rate of patients in critical sub-compartment 1 | 0.104 | 0.409 | 1/8 | 0.441 | |
| $rb_{1,2}$ | Proportionality coefficient of intensifications / relaxations between $b_1$ and $b_2$ | 0.379 | 9.18 | - | - | |
| $p_{S,M}$ | Fraction of unreported cases | 0.499 | 0.102 | 1/2 | | |
| $p_{crit}$ ($p_{crit,0}$) | Probability of becoming critical after developing symptoms (initial value) | 0.0765 | 0.706 | | - | |
| $p_{death}$ ($p_{death,0}$) | Probability of death after becoming critical (initial value) | 0.119 | 1.24 | | - | |
| $p_{death,S,0}$ | Proportionality coefficient for evaluating probability of death after developing symptoms without becoming critical, see (S2.3) | 0.587 | 8.04 | - | - | |

## Parameter values for Saxony

**Supplemental Table S6: Time points of changes in infectivity and respective values for Saxony.** We used fixed (known due to Governmental decisions or random events) and estimated time points of changes of NPI / contact behaviour and events and respective changes in infectivity of asymptomatic subjects. We provide estimates and relative standard errors of the infectivity starting with the date mentioned (3rd to 5th column).

| Numbers | Type of NPI/behaviour change | Estimated new infectivity | Relative standard error, % | Date | Source | Standard error (days) |
|---|---|---|---|---|---|---|
| 1 | Intensification | 0.606 | 0.877 | 10.03.2020 | Fixed | - |
| 2 | Intensification | 0.120 | 5.41 | 15.03.2020 | Fixed | - |
| 3 | Intensification | 0.0904 | 1.15 | 22.03.2020 | Fixed | - |
| 4 | Relaxation | 0.103 | 1.98 | 02.04.2020 | Estimated | 0.541 |
| 5 | Intensification | 0.0907 | 3.12 | 14.04.2020 | Estimated | 0.237 |
| 6 | Relaxation | 0.302 | 0.965 | 30.04.2020 | Fixed | - |
| 7 | Intensification | 0.0606 | 6.08 | 07.05.2020 | Fixed | - |
| 8 | Intensification | 0.0385 | 4.21 | 14.05.2020 | Fixed | - |
| 9 | Relaxation | 0.0601 | 0.199 | 19.05.2020 | Estimated | 0.487 |

| 10 | Relaxation | 0.817 | 0.505 | 04.06.2020 | Estimated | 0.603 |
| 11 | Intensification | 0.0344 | 4.18 | 11.06.2020 | Estimated | 0.456 |
| 12 | Relaxation | 0.219 | 3.23 | 30.06.2020 | Estimated | 0.298 |
| 13 | Intensification | 0.149 | 1.13 | 16.08.2020 | Estimated | 0.312 |
| 14 | Relaxation | 0.213 | 2.29 | 26.08.2020 | Estimated | 0.578 |
| 15 | Relaxation | 0.297 | 0.78 | 04.10.2020 | Estimated | 0.209 |
| 16 | Intensification | 0.185 | 1.26 | 21.10.2020 | Estimated | 0.352 |
| 17 | Intensification | 0.152 | 5.93 | 30.10.2020 | Fixed | - |
| 18 | Relaxation | 0.201 | 0.826 | 11.11.2020 | Estimated | 1.21 |
| 19 | Relaxation | 0.207 | 0.652 | 19.11.2020 | Estimated | 0.318 |
| 20 | Intensification | 0.201 | 2.13 | 22.11.2020 | Estimated | 0.554 |
| 21 | Intensification | 0.0672 | 1.87 | 10.12.2020 | Fixed | - |
| 22 | Relaxation | 0.228 | 1.36 | 18.12.2020 | Estimated | 0.426 |
| 23 | Intensification | 0.0937 | 5.09 | 01.01.2021 | Estimated | 0.141 |
| 24 | Relaxation | 0.120 | 9.78 | 14.01.2021 | Estimated | - |
| 25 | Relaxation | 0.229 | 10.1 | 05.02.2021 | Estimated | - |
| 26 | Intensification | 0.150 | 11.5 | 15.02.2021 | Estimated | - |
| 27 | Relaxation | 0.199 | 0.95 | 26.02.2021 | Estimated | - |
| 28 | Relaxation | 0.210 | 25.7 | 18.03.2021 | Fixed | - |

**Supplemental Table S7: Step functions of $p_{crit}$ and $p_{death}$ for Saxony.** We present estimates for the steps of the functions $p_{crit}$ and $p_{death}$ at the specified dates and respective standard errors for Saxony. We also provide the standard error of the estimated time point (last column).

| Parameter | Description | Estimate | Relative standard error, % | Date respective controls | Standard error (days) |
|---|---|---|---|---|---|
| $\alpha_{crit,1}$ | Relative values of $p_{crit}$ starting at the respective date | 2.15 | 0.98 | 24.03.2020 | 0.34 |
| $\alpha_{crit,2}$ | | 1.99 | 4.22 | 10.04.2020 | 0.672 |
| $\alpha_{crit,3}$ | | 1.01 | 3.54 | 11.05.2020 | 1.25 |
| $\alpha_{crit,4}$ | | 2.54 | 2.49 | 05.06.2020 | 3.73 |
| $\alpha_{crit,5}$ | | 1.50 | 1.26 | 02.07.2020 | 4.36 |
| $\alpha_{crit,6}$ | | 1.19 | 3.41 | 27.07.2020 | 0.75 |
| $\alpha_{crit,7}$ | | 0.764 | 0.478 | 29.08.2020 | 4.93 |
| $\alpha_{crit,8}$ | | 0.398 | 5.12 | 18.09.2020 | 2.96 |
| $\alpha_{crit,9}$ | | 0.300 | 2.09 | 25.09.2020 | 2.12 |
| $\alpha_{crit,10}$ | | 0.528 | 2.16 | 13.10.2020 | 0.49 |
| $\alpha_{crit,11}$ | | 0.908 | 3.72 | 26.10.2020 | 1.15 |
| $\alpha_{crit,12}$ | | 0.999 | 2.43 | 01.12.2020 | 5.31 |
| $\alpha_{crit,13}$ | | 1.76 | 1.91 | 26.12.2020 | 2.06 |
| $\alpha_{crit,14}$ | | 2.01 | 1.56 | 10.01.2021 | 0.67 |
| $\alpha_{crit,15}$ | | 2.99 | 0.98 | 25.01.2021 | 2.15 |
| $\alpha_{crit,16}$ | | 2.68 | 5.77 | 13.02.2021 | 4.12 |
| $\alpha_{crit,17}$ | | 1.11 | 1.33 | 05.03.2021 | 5.11 |
| $\alpha_{crit,18}$ | | 0.700 | 79.2 | 04.03.2021 | - |
| $\alpha_{death,1}$ | Relative values of $p_{death}$ starting at the respective date | 0.655 | 2.26 | 04.04.2020 | 0.241 |
| $\alpha_{death,2}$ | | 3.58 | 6.81 | 24.04.2020 | 0.335 |
| $\alpha_{death,3}$ | | 1.94 | 5.32 | 17.05.2020 | 1.01 |
| $\alpha_{death,4}$ | | 0.743 | 1.07 | 08.06.2020 | 0.619 |
| $\alpha_{death,5}$ | | 0.296 | 4.32 | 07.07.2020 | 5.60 |
| $\alpha_{death,6}$ | | 0.401 | 1.56 | 04.08.2020 | 6.13 |
| $\alpha_{death,7}$ | | 0.142 | 6.77 | 26.08.2020 | 4.43 |
| $\alpha_{death,8}$ | | 0.473 | 9.05 | 27.09.2020 | 0.95 |
| $\alpha_{death,9}$ | | 0.314 | 1.42 | 03.10.2020 | 1.27 |
| $\alpha_{death,10}$ | | 0.638 | 0.84 | 02.11.2020 | 3.62 |
| $\alpha_{death,11}$ | | 1.41 | 0.9 | 16.11.2020 | 1.19 |
| $\alpha_{death,12}$ | | 1.64 | 2.31 | 01.12.2020 | 1.63 |
| $\alpha_{death,13}$ | | 2.54 | 1.555 | 20.12.2020 | 0.903 |
| $\alpha_{death,14}$ | | 2.66 | 3.89 | 08.01.2021 | 5.52 |
| $\alpha_{death,15}$ | | 3.48 | 5.53 | 19.01.2021 | 1.08 |
| $\alpha_{death,16}$ | | 2.31 | 4.9 | 09.02.2021 | 0.383 |
| $\alpha_{death,17}$ | | 1.22 | 2.76 | 26.02.2021 | 2.06 |
| $\alpha_{death,18}$ | | 0.807 | 4.03 | 07.03.2021 | 5.34 |
| $\alpha_{death,19}$ | | 1.09 | 70.1 | 11.03.2021 | - |

**Supplemental Table S8: Parameter estimates and comparison with average priors for the parameter settings for Saxony.** We present estimated parameters of the SECIR model and initial conditions of control parameters and their respective standard errors for the parametrization of the epidemic in Saxony. We also perform a formal comparison of estimates and expected priors using t-Test.

| Parameter | Description | Posterior estimate | Relative standard error, % | Prior value | p-value |
|---|---|---|---|---|---|
| influx | Initial influx of infections into compartment $E$ until first interventions | 68.1 | 6.17 | - | - |
| $r_1$ | Infection rate through asymptomatic subjects | 1.61 | 1.32 | - | - |
| $r_3$ | Transit rate for compartment E (latent time) | 0.270 | 0.234 | 1/3 | 0.221 |
| $r_4$ | Transit rate for asymptomatic sub-compartments | 0.697 | 0.691 | 3/5 | 0.357 |
| $r_{4,b}$ | Rate of development of symptoms after infection | 0.294 | 3.27 | 1/2.5 | 0.489 |
| $r_5$ | Transit rate for symptomatic sub-compartments | 1.11 | 2.13 | 3/2.5 | 0.236 |
| $r_6$ | Rate of development of critical state after being symptomatic | 0.170 | 1.46 | 1/5 | 0.495 |
| $r_7$ | Transit rate for critical state sub-compartment | 0.198 | 0.659 | 3/17 | 0.372 |
| $r_8$ | Death rate of patients in critical sub-compartment 1 | 0.140 | 1.33 | 1/8 | 0.393 |
| $rb_{1,2}$ | Proportional coefficient of intensifications/relaxations between $b_1$ and $b_2$ | 0.248 | 15.5 | - | - |
| $p_{S,M}$ | Fraction of unreported cases | 0.509 | 5.37 | 1/2 | |
| $p_{crit}$ $(p_{crit,0})$ | Probability of becoming critical after developing symptoms (initial value) | 0.0794 | 1.76 | - | - |
| $p_{death}$ $(p_{death,0})$ | Probability of death after becoming critical (initial value) | 0.137 | 0.957 | - | - |
| $p_{death,S,0}$ | Proportionality coefficient for evaluating probability of death after developing symptoms without becoming critical, see (S2.3) | 0.719 | 7.3 | - | - |

References

1. Adiga A, Dubhashi D, Lewis B, Marathe M, Venkatramanan S, Vullikanti A. Mathematical Models for COVID-19 Pandemic: A Comparative Analysis. J Indian Inst Sci. 2020:1–15. Epub 2020/10/30. doi: 10.1007/s41745-020-00200-6 PMID: 33144763.
2. Flaxman S, Mishra S, Gandy A, Unwin HJT, Mellan TA, Coupland H, et al. Estimating the effects of non-pharmaceutical interventions on COVID-19 in Europe. Nature. 2020; 584:257–61. Epub 2020/06/08. doi: 10.1038/s41586-020-2405-7 PMID: 32512579.
3. Bo Y, Guo C, Lin C, Zeng Y, Li HB, Zhang Y, et al. Effectiveness of non-pharmaceutical interventions on COVID-19 transmission in 190 countries from 23 January to 13 April 2020. Int J Infect Dis. 2021; 102:247–53. Epub 2020/10/29. doi: 10.1016/j.ijid.2020.10.066 PMID: 33129965.
4. Harris JE. Overcoming Reporting Delays Is Critical to Timely Epidemic Monitoring: The Case of COVID-19 in New York City. ; 2020.
5. Böttcher S, Oh D-Y, Staat D, Stern D, Albrecht S, Wilrich N, et al. Erfassung der SARS-CoV-2-Testzahlen in Deutschland (Stand 2.12.2020). Robert Koch-Institut; 2020.

6. McCulloh I, Kiernan K, Kent T. Improved Estimation of Daily COVID-19 Rate from Incomplete Data. 2020 Fourth International Conference on Multimedia Computing, Networking and Applications (MCNA). IEEE; 19.10.2020 - 22.10.2020. pp. 153–8.
7. Georgatzis K, Williams CKI, Hawthorne C. Input-Output Non-Linear Dynamical Systems applied to Physiological Condition Monitoring. Proceedings of the 1st Machine Learning for Healthcare Conference 2016: PMLR; 19 August 2016 through 20 August 2016 [updated 2016 Aug 19 through 2016 Aug 20].
8. Wise J. Covid-19: New coronavirus variant is identified in UK. BMJ. 2020; 371:m4857. Epub 2020/12/16. doi: 10.1136/bmj.m4857 PMID: 33328153.
9. COVID-19. clinical aspects. Robert Koch Institute [updated 16 Sep 2021; cited 16 Sep 2021]. Available from: https://www.rki.de/DE/Content/InfAZ/N/Neuartiges_Coronavirus/Daten/Klinische_Aspekte.html.
10. Tagesreport-Archiv. Deutsche Interdisziplinäre Vereinigung für Intensiv- und Notfallmedizin (DIVI) e.V. [updated 29 Mar 2021; cited 29 Mar 2021]. Available from: https://www.divi.de/divi-intensivregister-tagesreport-archiv.
11. George Delgado, MD, FAAFP, John Safranek, MD, PhD, Bill Goyette, BS, MSEE, Richard Spady P. Reported versus Actual Date of Death. "Reported" versus "Actual": Two Different Things. Available from: https://covidplanningtools.com/reported-versus-actual-date-of-death/.
12. Bericht zu Virusvarianten von SARS-CoV-2 in Deutschland, insbesondere zur Variant of Concern (VOC) B.1.1.7. Robert Koch Institute [updated 3 Mar 2021; cited 3 Mar 2021]. Available from: https://www.rki.de/DE/Content/InfAZ/N/Neuartiges_Coronavirus/DESH/Bericht_VOC_2021-03-03.pdf?__blob=publicationFile.
13. Kheifetz Y, Scholz M. Modeling individual time courses of thrombopoiesis during multi-cyclic chemotherapy. PLoS Comput Biol. 2019; 15:e1006775. Epub 2019/03/06. doi: 10.1371/journal.pcbi.1006775 PMID: 30840616.
14. Kuhn E, Lavielle M. Coupling a stochastic approximation version of EM with an MCMC procedure. ESAIM: PS. 2004; 8:115–31. doi: 10.1051/ps:2004007.
15. Konstantinos Georgatzis, Christopher K.I. Williams, Christopher Hawthorne. Input-Output Non-Linear Dynamical Systems applied to PhysiologicalCondition Monitoring. ; 2016.
16. Meineke FA, Löbe M, Stäubert S. Introducing Technical Aspects of Research Data Management in the Leipzig Health Atlas. Stud Health Technol Inform. 2018; 247:426–30.
17. Hale T, Angrist N, Goldszmidt R, Kira B, Petherick A, Phillips T, et al. A global panel database of pandemic policies (Oxford COVID-19 Government Response Tracker). Nat Hum Behav. 2021; 5:529–38. Epub 2021/03/08. doi: 10.1038/s41562-021-01079-8 PMID: 33686204.
18. COVID-19 Government Response Tracker [updated 1 Aug 2021; cited 1 Aug 2021]. Available from: https://www.bsg.ox.ac.uk/research/research-projects/covid-19-government-response-tracker.
19. Mullen JL, Tsueng G, Latif AA, Alkuzweny M, Cano M, Haag E, et al. Outbreak.info. a standardized, open-source database of COVID-19 resources and epidemiology data [cited 18 Jul 2021]. Available from: https://outbreak.info.
20. Aktuelle Entwicklung der COVID-19 Epidemie in Leipzig und Sachsen. Bulletin 14 vom 20.02.2021. Available from: https://www.imise.uni-leipzig.de/sites/www.imise.uni-leipzig.de/files/files/uploads/Medien/bulletin_n14_covid19_sachsen__2021_02_22_v11.pdf.
21. Khailaie S, Mitra T, Bandyophadhyay A, Schips M, Mascheroni P, Vanella P, et al. Development of the reproduction number from coronavirus SARS-CoV-2 case data in Germany and implications for political measures. ; 2020.

22. Barbarossa MV, Fuhrmann J, Meinke JH, Krieg S, Varma HV, Castelletti N, et al. Modeling the spread of COVID-19 in Germany: Early assessment and possible scenarios. PLoS One. 2020; 15:e0238559. Epub 2020/09/04. doi: 10.1371/journal.pone.0238559 PMID: 32886696.
23. Der Heiden M an, Buchholz U. Modellierung von Beispielszenarien der SARS-CoV-2-Epidemie 2020 in Deutschland. Robert Koch-Institut; 2020.
24. Scholz S, Waize M, Weidemann F, Treskova-Schwarzbach M, Haas L, Harder T, et al. Einfluss von Impfungen und Kontaktreduktionen auf die dritte Welle der SARS-CoV-2-Pandemie und perspektivische Rückkehr zu prä-pandemischem Kontaktverhalten. 2021. doi: 10.25646/8256.
25. Kucharski AJ, Klepac P, Conlan AJK, Kissler SM, Tang ML, Fry H, et al. Effectiveness of isolation, testing, contact tracing, and physical distancing on reducing transmission of SARS-CoV-2 in different settings: a mathematical modelling study. Lancet Infect Dis. 2020; 20:1151–60. Epub 2020/06/16. doi: 10.1016/S1473-3099(20)30457-6 PMID: 32559451.
26. Quilty BJ, Clifford S, Hellewell J, Russell TW, Kucharski AJ, Flasche S, et al. Quarantine and testing strategies in contact tracing for SARS-CoV-2: a modelling study. Lancet Public Health. 2021; 6:e175-e183. Epub 2021/01/21. doi: 10.1016/S2468-2667(20)30308-X PMID: 33484644.
27. Bracher J, Wolffram D, Deuschel J, Görgen K, Ketterer JL, Ullrich A, et al. A pre-registered short-term forecasting study of COVID-19 in Germany and Poland during the second wave. Nat Commun. 2021; 12:5173. Epub 2021/08/27. doi: 10.1038/s41467-021-25207-0 PMID: 34453047.
28. Friberg LE, Henningsson A, Maas H, Nguyen L, Karlsson MO. Model of chemotherapy-induced myelosuppression with parameter consistency across drugs. J Clin Oncol. 2002; 20:4713–21. doi: 10.1200/JCO.2002.02.140 PMID: 12488418.
29. Maire F, Friel N, Mira A, Raftery AE. Adaptive Incremental Mixture Markov Chain Monte Carlo. J Comput Graph Stat. 2019; 28:790–805. Epub 2019/06/07. doi: 10.1080/10618600.2019.1598872 PMID: 32410811.
30. Dempster AP, Laird NM, Rubin DB. Maximum Likelihood from Incomplete Data Via the EM Algorithm. Journal of the Royal Statistical Society: Series B (Methodological). 1977; 39:1–22. doi: 10.1111/j.2517-6161.1977.tb01600.x.
31. Tierney L. Markov Chains for Exploring Posterior Distributions. Ann Statist. 1994; 22. doi: 10.1214/aos/1176325750.
32. Haario H, Saksman E, Tamminen J. An Adaptive Metropolis Algorithm. Bernoulli. 2001; 7:223. doi: 10.2307/3318737.
33. Stebegg M, Kumar SD, Silva-Cayetano A, Fonseca VR, Linterman MA, Graca L. Regulation of the Germinal Center Response. Front Immunol. 2018; 9:2469. Epub 2018/10/25. doi: 10.3389/fimmu.2018.02469 PMID: 30410492.
34. Wölfel R, Corman VM, Guggemos W, Seilmaier M, Zange S, Müller MA, et al. Virological assessment of hospitalized patients with COVID-2019. Nature. 2020; 581:465–9. Epub 2020/04/01. doi: 10.1038/s41586-020-2196-x PMID: 32235945.
35. Hu Z, Song C, Xu C, Jin G, Chen Y, Xu X, et al. Clinical characteristics of 24 asymptomatic infections with COVID-19 screened among close contacts in Nanjing, China. Sci China Life Sci. 2020; 63:706–11. Epub 2020/03/04. doi: 10.1007/s11427-020-1661-4 PMID: 32146694.
36. Li R, Pei S, Chen B, Song Y, Zhang T, Yang W, et al. Substantial undocumented infection facilitates the rapid dissemination of novel coronavirus (SARS-CoV-2). Science. 2020; 368:489–93. Epub 2020/03/16. doi: 10.1126/science.abb3221 PMID: 32179701.
37. Hao X, Cheng S, Wu D, Wu T, Lin X, Wang C. Reconstruction of the full transmission dynamics of COVID-19 in Wuhan. Nature. 2020; 584:420–4. Epub 2020/07/16. doi: 10.1038/s41586-020-2554-8 PMID: 32674112.

38. He X, Lau EHY, Wu P, Deng X, Wang J, Hao X, et al. Temporal dynamics in viral shedding and transmissibility of COVID-19. Nat Med. 2020; 26:672–5. Epub 2020/04/15.
doi: 10.1038/s41591-020-0869-5 PMID: 32296168.
39. Neuartiges_Coronavirus/Steckbrief. Robert Koch Institute [updated 1 Aug 2021; cited 1 Aug 2021]. Available from:
https://www.rki.de/DE/Content/InfAZ/N/Neuartiges_Coronavirus/Steckbrief.html;jsessionid=5BE662E8D7163EB2CE5E93F48148AF32.internet081?nn=13490888#doc13776792bodyText5.
40. Nishiura H, Linton NM, Akhmetzhanov AR. Serial interval of novel coronavirus (COVID-19) infections. Int J Infect Dis. 2020; 93:284–6. Epub 2020/03/04. doi: 10.1016/j.ijid.2020.02.060 PMID: 32145466.
41. Tindale LC, Stockdale JE, Coombe M, Garlock ES, Lau WYV, Saraswat M, et al. Evidence for transmission of COVID-19 prior to symptom onset. Elife. 2020; 9. Epub 2020/06/22.
doi: 10.7554/eLife.57149 PMID: 32568070.
42. Böhmer MM, Buchholz U, Corman VM, Hoch M, Katz K, Marosevic DV, et al. Investigation of a COVID-19 outbreak in Germany resulting from a single travel-associated primary case: a case series. The Lancet Infectious Diseases. 2020; 20:920–8. doi: 10.1016/S1473-3099(20)30314-5.
43. Ganyani T, Kremer C, Chen D, Torneri A, Faes C, Wallinga J, et al. Estimating the generation interval for coronavirus disease (COVID-19) based on symptom onset data, March 2020. Euro Surveill. 2020; 25. doi: 10.2807/1560-7917.ES.2020.25.17.2000257 PMID: 32372755.
44. Buitrago-Garcia D, Egli-Gany D, Counotte MJ, Hossmann S, Imeri H, Ipekci AM, et al. Occurrence and transmission potential of asymptomatic and presymptomatic SARS-CoV-2 infections: A living systematic review and meta-analysis. PLoS Med. 2020; 17:e1003346. Epub 2020/09/22.
doi: 10.1371/journal.pmed.1003346 PMID: 32960881.
45. Oran DP, Topol EJ. Prevalence of Asymptomatic SARS-CoV-2 Infection : A Narrative Review. Ann Intern Med. 2020; 173:362–7. Epub 2020/06/03. doi: 10.7326/M20-3012 PMID: 32491919.
46. Byambasuren O, Cardona M, Bell K, Clark J, McLaws M-L, Glasziou P. Estimating the extent of asymptomatic COVID-19 and its potential for community transmission: Systematic review and meta-analysis. Official Journal of the Association of Medical Microbiology and Infectious Disease Canada. 2020; 5:223–34. doi: 10.3138/jammi-2020-0030.
47. Zhou F, Yu T, Du R, Fan G, Liu Y, Liu Z, et al. Clinical course and risk factors for mortality of adult inpatients with COVID-19 in Wuhan, China: a retrospective cohort study. The Lancet. 2020; 395:1054–62. doi: 10.1016/S0140-6736(20)30566-3.
48. Sanche S, Lin YT, Xu C, Romero-Severson E, Hengartner N, Ke R. High Contagiousness and Rapid Spread of Severe Acute Respiratory Syndrome Coronavirus 2. Emerg Infect Dis. 2020; 26:1470–7. Epub 2020/06/21. doi: 10.3201/eid2607.200282 PMID: 32255761.
49. COVID-19 National Emergency Response Center. Coronavirus Disease-19: The First 7,755 Cases in the Republic of Korea. Osong Public Health Res Perspect. 2020; 11:85–90.
doi: 10.24171/j.phrp.2020.11.2.05 PMID: 32257774.
50. Wu Z, McGoogan JM. Characteristics of and Important Lessons From the Coronavirus Disease 2019 (COVID-19) Outbreak in China: Summary of a Report of 72 314 Cases From the Chinese Center for Disease Control and Prevention. JAMA. 2020; 323:1239–42.
doi: 10.1001/jama.2020.2648 PMID: 32091533.
51. Schuppert A, Theisen S, Fränkel P, Weber-Carstens S, Karagiannidis C. Bundesweites Belastungsmodell für Intensivstationen durch COVID-19. Med Klin Intensivmed Notfmed. 2021. Epub 2021/02/03. doi: 10.1007/s00063-021-00791-7 PMID: 33533980.
52. Tolksdorf K, Buda S, Schuler E, Wieler LH, Haas W. Eine höhere Letalität und lange Beatmungsdauer unterscheiden COVID-19 von schwer verlaufenden Atemwegsinfektionen in Grippewellen. 2020. doi: 10.25646/7111.

53. Karagiannidis C, Mostert C, Hentschker C, Voshaar T, Malzahn J, Schillinger G, et al. Case characteristics, resource use, and outcomes of 10 021 patients with COVID-19 admitted to 920 German hospitals: an observational study. The Lancet Respiratory Medicine. 2020; 8:853–62. doi: 10.1016/S2213-2600(20)30316-7.
54. Linton NM, Kobayashi T, Yang Y, Hayashi K, Akhmetzhanov AR, Jung S-M, et al. Incubation Period and Other Epidemiological Characteristics of 2019 Novel Coronavirus Infections with Right Truncation: A Statistical Analysis of Publicly Available Case Data. J Clin Med. 2020; 9. Epub 2020/02/17. doi: 10.3390/jcm9020538. PMID: 32079150.
55. Verity R, Okell LC, Dorigatti I, Winskill P, Whittaker C, Imai N, et al. Estimates of the severity of coronavirus disease 2019: a model-based analysis. The Lancet Infectious Diseases. 2020; 20:669–77. doi: 10.1016/S1473-3099(20)30243-7.
56. Gornyk D, Harries M, Glöckner S, Strengert M, Kerrinnes T, Bojara G, et al. SARS-CoV-2 seroprevalence in Germany - a population based sequential study in five regions. ; 2021.
57. Neuhauser H, Thamm R, Buttmann-Schweiger N, Fiebig J, Offergeld R, Poethko-Müller C, et al. Ergebnisse seroepidemiologischer Studien zu SARS-CoV-2 in Stichproben der Allgemeinbevölkerung und bei Blutspenderinnen und Blutspendern in Deutschland (Stand 03.12.2020). 2020. doi: 10.25646/7728.